\documentclass{article}

\usepackage{amsmath}
\usepackage{amsthm}
\usepackage{amsfonts}
\usepackage{graphicx}
\usepackage[utf8x]{inputenc}
\usepackage[T1]{fontenc}
\usepackage[margin=1.5in]{geometry}
\usepackage{bm}
\usepackage{lipsum}
\usepackage{xcolor}
\usepackage{bm}

\usepackage{natbib}

\graphicspath{{images/}}

\usepackage{pdfpages}

\usepackage{hyperref}
\usepackage[nobottomtitles*]{titlesec}

\usepackage[multiple, hang]{footmisc}
\usepackage{titlesec}

\usepackage{palatino}

\usepackage{fancyhdr}
\newcommand\blfootnote[1]{%
  \begingroup
  \renewcommand\thefootnote{}\footnote{#1}%
  \addtocounter{footnote}{-1}%
  \endgroup
}

\newcommand{\myparagraph}[1]{\paragraph{#1}\mbox{}\\}

\newcommand*{\captionsource}[2]{%
  \caption[{#1}]{%
    #1%
    \\\hspace{\linewidth}%
    \textit{Source:} #2%
    \\\hspace{0.5\linewidth}%
  }%
}

\newtheorem*{remark}{Remark}

\theoremstyle{definition}
\newtheorem{definition}{Definition}

\title{Harnessing value from data science in business: ensuring explainability and fairness of solutions}
\author{
    Krzysztof Chomiak\thanks{London School of Economics alumnus, \url{https://www.linkedin.com/in/krzysztof-chomiak/}}
    \and 
    Michał Miktus\thanks{Paris School of Economics alumnus, \url{https://www.linkedin.com/in/michalmiktus/}}
}

\begin{document}

\maketitle
% Here is the abstract.
\begin{abstract}
The paper introduces concepts of fairness and explainability (XAI) in artificial intelligence,  oriented to solve a sophisticated business problems. For fairness,  the authors discuss the bias-inducing specifics, as well as relevant mitigation
methods, concluding with a set of recipes for introducing fairness in data-driven organizations.
Additionally, for XAI, the authors audit specific algorithms paired with demonstrational business use-cases, discuss a plethora of techniques of
 explanations quality quantification  and provide an overview of future research avenues.
\end{abstract}
%----
\blfootnote{\newline The authors would like to thank Przemysław Biecek, Tomasz Trzciński and Tomasz Bartkowiak for all their careful, constructive and insightful comments in relation to this work.}

\tableofcontents

\section{Introduction}
Artificial Intelligence (AI) is incessantly flourishing in sophistication and complexity.
The latest developments of intelligent systems, trained to take autonomous and consistent decisions, lead to several transformational opportunities for business and society.
Consequently, machine learning has encroached into a plethora of business domains, many of them requiring high level of accountability, and thus transparency and fairness.

Undoubtedly, the full understanding and fairness of model structure and its outcomes is a prerequisite for building trust and adoption of AI systems.
It is particularly true in high-stakes disciplines requiring reliability and safety, such as healthcare or automated transportation, as well as in demanding industrial applications with critical economic implications, not to mention exploration of natural resources or climate change modelling.
Nevertheless, recent mainstream machine learning research focused primarily on improving accuracy scores on benchmark datasets, neglecting the non-trivial implications of relying on such models.

A good illustration of the bias embedded in contemporary machine learning algorithms is a widely-cited
technique proposed by \cite{ribeiro_why_2016} that could recognize wolves from husky dogs, reaching its
decision by considering solely the presence of snow in the background.
Another example with serious business implications is the controversy surrounding the black-box-based credit
scoring process behind the Apple Card with alleged discrimination based on gender.
\footnote{Example article covering the situation: \url{https://edition.cnn.com/2019/11/10/business/goldman-sachs-apple-card-discrimination/index.html}} 
Other instances include a predictive model for risk of repeating criminal offences with a strong ethnic bias
(Angwin et al., 2016), or finally the adversarial examples, where some attackers could selectively add
distortions to images to impose on an otherwise robust classifier to always choose a certain class \citep{szegedy2014intriguing}.  With an incessant intensification of instances of artificial intelligence missteps, the demand for ensuring fairness of models in a systematic manner is sprouting.

The above-mentioned shortcomings and obvious failures of machine learning methods led to the birth of research of fair and explainable AI, considered as  subfields of AI that aim to satisfy the inherent need to control and explain machine decisions, simultaneously guaranteeing that it will continue to perform as expected when deployed in a real-world environment.

The remainder of this paper is organized into two main sections:
\begin{itemize}
    \item Fairness in AI - the section discusses various definitions of fairness,
    analyzes sources of bias, provides methods of mitigating bias at different modelling phases,
    introduces state-of-the-art tools for fairness and
    finally discusses a high-level guideline of implementing AI fairness in an organization,
    \item eXplainable AI (XAI) - the section starts with an outline of fundamental reasons for XAI emergence.
    Then, basic characteristics of proper explainability methods are summarized,
    leading to the synopsis of XAI techniques applied before (pre-modelling explainability),
    during (explainable modelling), and after (post-modelling explainability) the modelling stage,
    with a comparison of their respective strengths and weaknesses,
    as well as an illustration of business use-cases employing explainable artificial intelligence algorithms.
    The final subsection provides measures for evaluating performance of models in explainability
    tasks, such as Lipschitz value or faithfulness indicator.
\end{itemize}

\section{Fairness in AI}

The fairness of AI algorithms is a burgeoning field of research that stems from the general requisite for decisions to be unfettered from bias and discrimination. In general, it can be delineated as \textit{the absence of any prejudice or favouritism toward an individual or a group based on their inherent or acquired characteristics}. Nevertheless, it has to be remembered that it is a highly nuanced concept and its definition differs across cultures, societies, and even over time. 

\begin{remark}
Statisticians proposed a plethora of mathematical criteria to be satisfied for a classifier or a model to be characterised as \textit{fair}. The non-exhaustive list of twenty-one fairness definitions can be found at the Arvind Narayanan tutorial at ACM FAT (Fairness, Accountability and Transparency) Conference in 2018. \footnote{\url{https://www.youtube.com/watch?v=jIXIuYdnyyk}}
\end{remark}

It has to be emphasised that eliminating AI bias is not just an ethical obligation, yet it should be regarded as an economic and competitive imperative. By imposing the AI systems to be fair, organizations may become more profitable and productive. Consider the market for credit, which has a distasteful history of redlining against racial minorities and consequently, to a large extent, effectively restraining them out of the financial system and limiting market size for credit providers. As part of its fairness governance policy, the credit bureau Experian refined its Boost program to permit clients with limited credit history to augment their scores with TV, phone bills or even Netflix history \footnote{\url{https://www.experian.com/consumer-products/score-boost.html}}. Since the program launch in January 2019, more than 4 million customers have participated, while more than 150 thousand people with weak credit rating have been able to be upgraded to a mediocre one, providing significantly more funding opportunities.

\subsection{Mathematical definition of fairness}

As mentioned in the previous section, it is relatively strenuous to align on one clear concept of fairness in AI. Most definitions of unbiasedness are connected to the group fairness, which handles with statistical fairness across the whole population. Alternatively, the complementary notion of individual fairness mandates that similar individuals should be treated equivalently regardless of their group membership.

In order to grasp the arduousness of exact fairness definition, please find below three main notions of group fairness. Consider input data, denoted by $\bm{x}$, and machine learning model $f(\bm{x})$ that aims to predict a binary outcome $\bm{y} \in \{0, 1\}$\footnote{Proposed definitions can be easily generalised for multilevel outcome.}. Please note that the outcomes may differ statistically between different populations. Moreover, assume that each record $\bm{x}$ possesses a binary \textit{protected attribute} $\bm{p}\in \{0, 1\}$, such as gender or ethnicity. In the literature observations where $\bm{p}=0$ are referred to as the \textit{deprived population}, and conversely instances where $\bm{p}=1$ as the \textit{favoured population}. Similarly, $\bm{y}=1$ denotes the favoured outcome (representing more desirable of the two possible results) and the goal is to estimate $\bm{\hat{y}}$ 

\subsubsection{Demographic parity}

Demographic parity or statistical parity claims that a predictor is \textbf{unbiased} if the prediction $\bm{\hat{y}}$ is independent of the protected attribute $\bm{p}$. In other words:
\[
\mathbf{P}(\bm{\hat{y}} | \bm{p}) = \mathbf{P}(\bm{\hat{y}})
\]

Because the identical proportion of each population is classified as positive, it may lead to different false positive and true positive rates if the true outcome $\bm{y}$ does actually vary with the protected attribute $\bm{p}$.

Consequently, the deviations from fairness may be measured by the statistical parity difference:
\[
\text{SPD} = \mathbf{P}(\bm{\hat{y}} = 1 | \bm{p} = 1) - \mathbf{P}(\bm{\hat{y}} = 1 | \bm{p} = 0)
\]

or equivalently the statistical parity ratio:
\[
\text{SPR} = \frac{\mathbf{P}(\bm{\hat{y}} = 1 | \bm{p} = 1)}{\mathbf{P}(\bm{\hat{y}} = 1 | \bm{p} = 0)}
\]

\subsubsection{Equality of odds}

Equality of odds requires that the prediction $\bm{\hat{y}}$ is conditionally independent to the protected attribute $\bm{p}$, given the true value $\bm{y}$. To put it differently:
\[
\mathbf{P}(\bm{\hat{y}} | \bm{y}, \bm{p}) = \mathbf{P}(\bm{\hat{y}} | \bm{y})
\]
 
As a consequence, the true positive rate and false positive rate are the same for each population.

\subsubsection{Equality of opportunity}

Equality of opportunity is the equality of odds focuses solely for positive outcome, namely:
\[
\mathbf{P}(\bm{\hat{y}} | \bm{y} = 1, \bm{p}) = \mathbf{P}(\bm{\hat{y}} | \bm{y} = 1)
\]

Similarly to the statistical parity difference, the equal opportunity difference is defined as:
\[
\text{EOD} = \mathbf{P}(\bm{\hat{y}} = 1, \bm{y} = 1, \bm{p} = 1) - \mathbf{P}(\bm{\hat{y}} = 1, \bm{y} = 1, \bm{p} = 0) 
\]

while the equal opportunity ratio:
\[
\text{EOR} = \frac{\mathbf{P}(\bm{\hat{y}} = 1, \bm{y} = 1, \bm{p} = 1)}{\mathbf{P}(\bm{\hat{y}} = 1, \bm{y} = 1, \bm{p} = 0)}
\]

\subsection{Sources of bias}

In order to fully unleash the profitable opportunities of AI, one needs to identify and understand the sources of discrimination. Biased human judgements can affect AI systems in two alternative ways: input data or employed algorithms.

\subsubsection{Input data bias}

First source of AI bias is inherently connected to the quality of input data. More specifically, the training dataset can be non-representative or it may possess the societal bias.

Machine learning models rely on large datasets in order to learn to identify patterns. As a consequence, infrequent motifs may be down-weighted in the sake of generalization abilities, leading the minority records to be unfairly neglected. Moreover, the lack of quality data may not be driven purely by the group's negligible size. Namely, the data collection methodology can exclude or disadvantage some classes, for instance if it is performed solely in one language. In addition, missing values leading to the records removal may be more prevalent in some factions than others.

For instance, \cite{pmlr-v81-buolamwini18a} evaluated the most commonly employed facial analysis algorithms and recognized the considerable disparities in their accuracy for classifying the darker-skinned females (error rates of up to 34.7\%) and the lighter-skinned males (0.8\%). The thorough investigation revealed that the training datasets did not contain enough examples of darker-skinned females for the model to perform satisfactorily well.

Along the same vein, even if the amount of data is sufficient to adequately represent each class, the training dataset may reflect existing societal and historical prejudices, such as the phenomenon that female workers are on average paid less. Such historical unfairness in data is often referred to as \textit{negative legacy}.

As an example, consider the natural language processing algorithm to predict the next word in the following sentence: \textit{father is to a doctor as mother is to a}. Even if it had been trained on the whole data from Wikipedia, due to the indelible historical and cultural gender stereotypes, it is most likely to finish with \textit{nurse}.

Lastly, the usage of proxies introduces the risk of skewing the data. The notable example of hospitals that chose \textit{healthcare spending} as a proxy for \textit{health} (in lieu of difficult-to-obtain medical records) without accounting for different healthcare spending patterns and unequal access to care means for Black and White patients affected millions of Americans \citep{obermeyer_dissecting_2019}. The proper input data increased the percentage of Black patients receiving additional help from 17.7\% to 46.5\%.

\subsubsection{Algorithmic bias}

A second source of bias arises due to the fact that machine learning algorithms generally reckon on correlation, rather than causal relationships. For example, it is relatively straightforward to design a model to confirm the positive relationship between the temperature levels and the proportion of female at managerial positions.

Furthermore, the model architecture may be more suitable for some classes than others. For instance, a linear model may be appropriate for one cohort but not for another.

What is more, the bias could be embedded in the choice of cost function, for example by restricting the set of possible decisions that the machine may make. Instead of a \textit{yes or no} resolution, it might be a \textit{one-out-of-many} verdict, introducing bias if a reasonable option to consider is eliminated.

\subsubsection{Detecting bias}

Understanding bias sources helps in their actual identification. Consequently, the following four steps are useful in bias detection:

\begin{enumerate}
    \item Verify whether all data groups have an identical probability of being assigned to the favourable outcome for protected attribute.
    \item Validate whether all factions of a protected attribute have corresponding positive predictive value.
    \item Authenticate that all classes of a protected attribute have predictive equality for false positive and false negative rates.
    \item Certify that statistical parity and equal opportunity ratios/differences are below certain level.
\end{enumerate}

\subsection{Mitigating bias}

Once the bias is detected, its eradication should be performed gradually at each stage of data science life cycle. 

Thus, before modelling, one can boost the fairness of the delivered solution by ensuring equal proportional representation across all groups. During modelling, one should focus on guaranteeing that model performance is fair for all groups for one or more sensitive features, for example by ensuring equivalent AUC metric for male and female instances. Finally, after modelling, one should assure that predictions and model output sustain an equal probability of false positive rate (or any other model performance metrics) for all classes with all protected features.

\begin{remark}
The choice of an appropriate fairness-boosting method should be grounded in an understanding of the source of bias and which fairness definition we are aiming for. For example, some of the more rigid methods described below aim at achieving \textit{demographic parity}, which should be used only together with a strong conviction and evidence that indeed the protected attribute does not influence the target variable.
\end{remark}

\subsubsection{Ante-hoc fairness}

The ante-hoc methods for ensuring fairness focus on modifying the data used for modelling in a way that makes the resulting model fair.

One of the simplest ways would be deleting the protected attributes. However, it has to be emphasised that ostracising an individual's protected attributes is not sufficient to assure the equity of the solution, because it may be reconstructed through subtle correlations in the input data.

\begin{remark}
The degree to which there are dependencies between the data $\bm{x}$ and the protected attribute $\bm{p}$ can be quantified using the \textit{latent prejudice} measure \citep{inproceedings_kamishima}, defined as:

\[
\text{LP} = \sum\limits_{\bm{x}, \bm{p}} \mathbf{P}(\bm{x} | \bm{p})\log\left[ \frac{\mathbf{P}(\bm{x} | \bm{p})}{\mathbf{P}(\bm{x}) \mathbf{P}(\bm{p})} \right]
\]

As the aforementioned measure rises, the protected attribute becomes more predictable from the data. Similarly, one can introduce the \textit{indirect prejudice} measure \citep{inproceedings_kamishima} to account for the dependencies between the labels $\bm{y}$ and the protected attribute $\bm{p}$:

\[
\text{IP} = \sum\limits_{\bm{y}, \bm{p}} \mathbf{P}(\bm{y} | \bm{p})\log\left[ \frac{\mathbf{P}(\bm{y} | \bm{p})}{\mathbf{P}(\bm{y}) \mathbf{P}(\bm{p})} \right]
\]

Theoretically, if one cannot predict the labels from the protected attribute and vice-versa, then there is no possibility for bias. 

\end{remark}

Ensuring fairness of to-be-delivered solution in a systematic way initiates at the data collection step, through the identification of lack of examples or groups and further data enhancement \citep{chen2018classifier}. It can be then closely followed by the modification of labels and/or input data, as well as the corresponding re-sampling and weighting through statistical calibration.

\textbf{Manipulating labels:} As a potential solution of fairness ensuring before modeling phase, \cite{kamiran_data_2012},  as an extension of their prior work \cite{4909197}, urged to modify some of the training labels. Their so-called \textit{massaging} approach initiates with the computation of the classifier on the original dataset in order to identify observations with predictions closest to the decision surface (in order to minimize the accuracy loss). Then, their labels are swapped (maintaining the overall class distribution) and the model is re-trained, ensuring that the disadvantaged group is more likely to occur. 

\textbf{Manipulating observed data:} Alternatively, \cite{feldman_certifying_2015} introduced \textit{disparate impact removal} technique to de-bias the dataset by mapping each of the the cumulative distributions $F_{0}(\bm{x}) = F(\bm{x} | \bm{p} = 0)$ and $F_{1}(\bm{x}) = F(\bm{x} | \bm{p} = 1)$ to a median distribution. In other words, it adapts each attribute to ensure the equality of the conditional marginal distributions based on the subsets of protected attribute with a given sensitive value. The main flaw of the proposed approach is that it treats each feature separately, thus ignoring the potential interactions. In addition, since the repair process outlined above is likely to degrade model's ability to classify accurately, one might consider a partially repaired data set instead, as a trade-off between the predictive power and the fairness. It can be achieved by simply moving each distribution only part way (for instance a quartile) towards the median distribution. 

\textbf{Manipulating labels and observed data:} \cite{calmon_data_2018} recommended to manipulate both the labels and the input data in order to probabilistically transform data pairs ${\bm{x}, \bm{y}}$ to new data values ${\bm{x'}, \bm{y'}}$ in a way that depends explicitly on the protected attribute $\bm{p}$. The aforementioned transformation is formulated as a utility change minimization problem with group fairness, individual distortion, and data fidelity constraints. Unlike previous techniques, it takes into consideration interactions between all of the data dimensions. Furthermore, it also enables an explicit control of individual fairness and the feasibility of multivariate, non-binary protected features. Nevertheless, since it is formulated as a probability table, it is mainly applicable for datasets with limited number of discrete input and output variables.

\textbf{Re-weighting data pairs:} If the input dataset contains unjustified dependencies between some data attributes and the class labels, \cite{4909197} suggested to balance the dataset such that all the cases where the protected attribute $\bm{p}$ predicts that the disadvantaged group will get a positive outcome obtain higher weights.

\textbf{Learning fair representations:} \cite{zemel2013} proposed a learning algorithm for fair classification that aimed to achieve both group and individual fairness. They formulated it as an optimization problem of finding a proper representation of the data with two clashing objectives: to encode the data as accurately as possible, while simultaneously befuddling any information about protected attributes. Since the above-mentioned mapping is determined during training, their approach could be classified either as a pre-processing technique or an in-processing algorithm. Moreover, modern research focuses on employing a variational fair autoencoder, based on deep variational autoencoders, as in \cite{louizos2017variational} or \cite{li_learning_2014}.

\subsubsection{In-processing fairness}

\textbf{Regularize for fairness}: Adding a discrimination-aware regularizer (a mathematical constraint to ensure fairness in the model) to existing ML algorithms as a part of the learning objective, for example in the form of mutual information between the model's prediction and the protected attribute \citep{inproceedings_kamishima}.

\textbf{Constrain to be meta-fair}: Regularization with standard fairness constraints is often non-convex and hard to be satisfied directly. Consequently, one can re-express fairness constraints and introduce some convex relaxation for the purpose of optimization. One possibility is to define unfairness as the covariance between the protected attribute and distances of the related feature vectors from the decision boundary and consequently minimize loss the function under fairness constraints or, interchangeably, to maximize fairness under accuracy constraints \citep{zafar2017fairness}.

\textbf{Adversarial de-biasing}: Drawing from research in reinforcement learning, another approach is to construct two models: one for classification task and the other for fairness assurance. To be more specific, one needs to maximize the classifier's prediction accuracy, while simultaneously minimize its adversary's competence to ascertain the protected attribute directly from the predictions. \cite{beutel_data_2017} proposed a typical adversarial infrastructure, obliging both classifiers to use a shared representation, while \cite{zhang_mitigating_2018} proffered to urge the adversary model to predict the protected attribute $\bm{p}$ from:

\begin{enumerate}
    \item the final classification to protect the demographic parity
    \item the final classification and the true class to guarantee the equality of odds
    \item the final classification and the true value for just one class to provide the equality of opportunity
\end{enumerate}

\textbf{Use surrogate models}: Alternatively, one can wrap a fair algorithm around baseline ML algorithms already in use. \cite{calders_three_2010} proposed a Two Naive Bayes technique, which trains distinct models for each protected attribute and iteratively modifies the observed probabilities towards the selected fairness metrics, based on the fairness of the combined model. A more sophisticated approach models the actual class labels by treating them as a latent variable, trying to predict the labels that the dataset would have if it had been free of discrimination from the very beginning \citep{calders_three_2010}.

\subsubsection{Post-hoc fairness}

It is difficult to remove bias once the classifier has already been trained, even for very simple cases \citep{DBLP:journals/corr/abs-1709-02012}. As a potential remedy, one can attempt to calibrate the prediction probability threshold to maintain fair outcomes for all groups with protected and sensitive features, yet at the cost of decreased accuracy.

\textbf{Discrimination-Aware Decision Tree Relabelling}: \cite{4909197} designed a strategy to modify the labels of leaves in a decision tree after training in order to satisfy fairness constraints. In other words, their approach aims to find the minimal subset of leaves which labels need to be flipped in order to obtain a less discriminative prediction. 

\textbf{Reject-Option-Based Classification}: This approach modifies the predictions of samples close to the decision boundary, usually near 0.5 in binary classification for balanced target variables, by incorporating a rejection option \citep{10754/564630}. To be more specific, if the prediction is contained in the critical region, all unprivileged individuals receive the favorable outcome, while for the predictions outside the critical region, they remain unchanged.

\subsection{Tools for fairness}

Since fairness in AI is incessantly gaining momentum, technology leaders propose a variety of tools to automate the fairness-establishment process. The list below is a compilation of most commonly used tools:

\begin{enumerate}
    \item What-If Tool by Google\footnote{Available at: \url{www.pair-code.github.io/what-if-tool/}}
    
    The What-if Tool is an open-source visualization tool that allows users to analyse an ML model. Currently, it supports both binary and multi-class classification, as well as the regression problems, permitting users to explore the effects of alternative thresholds and fairness criteria. Moreover, it envisions prediction counterfactual, by contrasting an observation to the most similar point where the model predicts a different outcome. 
    
     \item Fairlearn by Microsoft\footnote{Available at: \url{https://fairlearn.org}}
     
     Fairlearn is an open-source library to assess and ameliorate the fairness of AI systems. It is composed of two primary components: an interactive visualisation dashboard with a set of metric to help quantify the fairness of a model, as well as several algorithms that can be used to mitigate models discrimination (training or post-processing), for instance a method that applies relevant constraints in order to reduce a fair classification problem to a sequence of cost-sensitive classification ones.

    \item AI Fairness 360 by IBM\footnote{Available at: \url{www.aif360.mybluemix.net}}
    
    The AI Fairness 360 toolkit is an extensible open-source library, available for both R and Python, containing a comprehensive set of metrics for datasets and models to be validated for biases and algorithms to mitigate bias in machine learning models throughout the AI application life-cycle. It is tailored to translate academia research into the actual practice of domains as wide-ranging as finance, human capital management, healthcare, and education.
    
     \item Responsible AI Toolkit by PwC\footnote{Available at: \url{www.pwc.com/gx/en/issues/data-and-analytics/artificial-intelligence/what-is-responsible-ai.html}}
     
     Responsible AI Toolkit is a suite of tools designed to help to tame the power of AI ethically and responsibly. Due to being developed with various stakeholders such as regulators or board members, it aids to address the regulatory and compliance facets of AI-oriented businesses. 
     
     \item LinkedIn Fairness Toolkit (LiFT) by LinkedIn\footnote{Available at: \url{www.github.com/linkedin/LiFT}}
     
     The LinkedIn Fairness Toolkit (LiFT) is a Scala/Spark library that empowers the quantification of fairness in big data machine learning workflows, being feasible to be deployed in both training and scoring phases.
     
\end{enumerate}

\subsection{Putting fairness to work}

Fairness in AI is a rapidly evolving field and therefore it has not established yet the standard best practices for implementation.
However, it is already possible for an organization to benefit from thinking about fairness in AI applications.
To summarize the chapter on fairness,  an overview of specific points to consider in practice is provided:
\begin{itemize}
    \item Analyse whether the context in which one aims to utilize AI involves the potential for bias. 
    If it does, then AI can have one of two potential consequences: mitigating or reinforcing bias. Unfortunately, without proper introduction of fairness methods described in the previous sections, the most likely outcome
    is bias reinforcement. 
    \item Implement new ways of working, designed specifically to utilize the fairness tools and techniques.
    To ensure that an organization systematically considers issues of fairness, one needs to complete three to four initiatives:
    \begin{enumerate}
        \item Make fairness in AI an important topic on agenda, so that people across the
        the organization understand that they need to start treating it as an important business
        objective. This is in contrast to simply forcing a procedure or policy, which would likely
        have very limited real impact.
        \item Provide analytical teams with the tools needed to tackle fairness issues:
        the required knowledge (e.g., via specific training sessions) and 
        \textit{literally} tools, which
        in some cases might mean using open-source libraries, but in some more restrictive environments
        it might require an investment to either buy or develop some tooling in-house.
        \item Find opportunities for operational changes that will aid the technical work on de-biasing. For example, it might be possible to change data collection processes, so that future model training will require less algorithmic tweaking for fairness.
        \item For \textit{safety-critical} areas\footnote{Areas in which utilizing biased AI
        could have serious negative consequences for the organization, most commonly from a regulatory
        point of view, but in some cases also from limiting the addressable market or causing serious
        harm to the company's image.}, create an additional audit function, which would 
        independently validate if the fairness objectives have been achieved.
    \end{enumerate}
    \item Extend the scope of work on fairness from AI to other business areas. For example, one
    might uncover some unfair processes when models trained on decision made in the past by humans
    show signs of bias.
    \item Identify business areas which can be handled directly by AI models and ones which require
    a \textit{human-in-the-loop}. In some cases (e.g., previously mentioned safety or low model
    quality) it makes sense to add a human as the final decision maker, to double check for fairness.
    It is tightly connected with severe negative cost implications, so should not be treated as a default method
    of de-biasing,
    \item Engage in a broader discussion, outside an organization.
    As fairness in AI is the current \textit{cutting-edge} research, there are a lot of
    opportunities to shape the future standards of it.  Undoubtedly, it will be achieved significantly
    faster with support from practitioners. Consequently, some of the ways of contributing might be providing
    researchers with access to datasets or sending teams to present their work at
    conferences.
\end{itemize}

\section{eXplainable Artificial Intelligence}

The second chapter of this article treats the topic of explainability in AI. 
XAI is presented after fairness for two reasons: 
\begin{enumerate}
    \item the future of XAI expands into a notion of Responsible AI,
    which provides not only an explainability with a satisfactory level of performance, 
    but excavate the concept with the apprehension of fairness, accountability,
    privacy and data fusion \citep{arrieta_explainable_2019},
    \item research in XAI is more advanced than in AI fairness, which translates to more 
    methods, which the reader might find applicable for their fairness use-case.
\end{enumerate}

As this chapter puts a great weight on specific XAI
techniques, we provide business
use-cases for the most relevant ones.

\subsection{Rationalization of XAI}
To start with, cooperation between agents, such as algorithms and humans, is based intensely on trust.
Improving transparency can diminish the inherent reluctance towards the algorithmic prescriptions,
since stakeholders can comprehend all the reasoning of each decision taken.
XAI provides a sense of control and safety, as the infrastructure owner clearly understands
the levers of its AI system's behaviour, which in turn can be subjected to a safety guidelines
and alert on their violation.

Moreover, the ability to explain a data-driven decision based on machine learning is particularly
important in banks, insurance companies, healthcare providers and other highly regulated industries.
They are prone to legal requirements, which oblige them to explain the rationale under every single
decision. Additionally, they may be constrained by the laws against discrimination, demanding the
ethics-related justifications (so called AI fairness), replacing \textit{black box} by \textit{glass
box}\footnote{A system that audits both the inputs and the outputs of a model, with the objective of
verifying its adherence to ethical and socio-legal values, thus developing value-based explanations.
For the more detailed distinction between black box, white box and glass box please refer to \cite{10.1007/11763864_18}.} In other words, interpretability helps ensure impartiality in decision-making,
allowing to detect, and consequently, correct from bias in the training dataset.

Other applications of XAI are knowledge extraction, permitting to learn new, robust insights from the
model and helping to pinpoint deficiencies both in data and feature behaviours. Moreover, the
interpretation does not have to be a good explanation per-se, but only guide the user to potential
flaws. Explainable artificial intelligence can help in debugging the mis-predictions by virtues of
expert knowledge and experience, ergo boosting the model’s performance. Furthermore, it can lead to
use of a model as a starting point for causal research, resulting in transferability of learning.
The mere understanding of the inner relations taking place within a model facilitates the ability
of a user to reuse this knowledge in another problem.

\begin{remark}
There are noteworthy differences among the concepts of explainability and interpretability. Namely, interpretability invokes a passive observation of the decision and judging its logic with human common sense. Reciprocally, explainability can be regarded as an active characteristic of a model, requiring interaction in order to clarify or detail its internal functions. For the sake of transparency, the notions of explainability, interpretability, comprehensibility and transparency will be henceforth used interchangeably.
\end{remark}

\subsection{Desired characteristics of XAI}
The field of data science is inherently linked to the performance versus explainability conundrum. One approach to achieve explainability in AI systems is to implement machine learning algorithms that are fundamentally explainable. For instance, classical/traditional methods of machine learning, such as decision trees or Bayesian classifiers, sometimes can relatively cheaply provide the traceability required for decisive AI systems without sacrificing \textbf{too much} performance. Conversely, more sophisticated and thus potentially more powerful algorithms such as neural networks or ensemble methods, like random forests, cede transparency and explainability for accuracy. Since XAI is designed to reconcile the feuding trade-offs, it has to possess five main characteristics:
\begin{enumerate}
    \item \textit{Global perspective}: XAI is required to present a coherent picture, generalizing local to global understanding and supporting a well-defined decision-making task;
    \item \textit{Interpretability}: XAI needs to provide the easily understandable results, highlighting only a few, influential features;
    \item \textit{Local fidelity}: XAI should be locally faithful, meaning reflecting the overall model performance (also referred to as accuracy);
    \item \textit{Model-agnostic property}: XAI (particularly post-hoc) should not depend on the underlying technique used;
    \item \textit{Efficiency}: XAI should achieve understandability in a reasonable amount of time;
\end{enumerate}

On the other hand, XAI techniques differ significantly among themselves as far as the focus of the explanation is concerned. Before selecting the proper algorithm, one needs to answer the following questions:

\myparagraph{At what stage of the ecosystem is explainability needed?}
There are three principal set of routines used to develop explainable systems: ante-hoc, contemporaneous and post-hoc. Ante-hoc techniques (often referred to as pre-modelling explainability) entail incorporating explainability into a model from the beginning. Contemporaneous method (explainable modelling) avoids the black-box problem from the outset by developing a model that is explainable by design. Finally, post-hoc procedures (post-modelling explainability) permit models to be trained unintermittedly, with explainability being incorporated at the testing time. 

\myparagraph{Who is the designator of explainability?}
The selected algorithm of XAI should reflect the role-specific goal of end user. The artificial intelligence experts usually require a mechanic explanation, for instance a sensitivity of a given layer of neural network to input data, extremely useful in debugging and validating a model. On the other hand, a business analyst tends to necessitate a functional explanation, for example a main driver behind the machine prediction, advantageous in checking the legal compliance of a model.

\myparagraph{What is the scope of explainability to be achieved?}
The microeconomic studies are apt to favour an instance prediction, often referred to as local explanations, which provide a reasoning for a specific instance of a class, such as the arguments for the unfavourable credit decision for an agent. On the other hand, the macroeconomic research is likely to focus on all model prediction, also named as global explanations, due to their relative simplicity to be translated into a policy advice.

\myparagraph{What level of complexity should be realized?}
The choice of the explanation family should mirror the desired complexity of an explanation. Despite the fact that the XAI algorithms should be model-agnostic, mainly due to the computational complexity some of them perform better in the case of a single output, like a prediction of a classifier, while some are more suited for the sequence explanation, such as Long Short-Term Memory (LSTM) neural networks \citep{sak_2014}. 

\subsection{XAI techniques}
The following section depicts a non-exhaustive overview of some of the most universal XAI methodologies and approaches, split into three main groups: ante-hoc, contemporaneous and post-hoc techniques.

\subsubsection{Ante-hoc explainability}
The behaviour of a plethora of AI models is, to a large extent, driven by datasets employed in their training stage. Hence, the main objective of pre-modelling explainability is to fully understand and construe the data. Ante-hoc methods can be partitioned into four main categories: dataset standardization (including its description regularity), exploratory data analysis, dataset squashing and explainable feature engineering.

\myparagraph{Dataset standardization}
Dataset standardization is a data processing workflow that aims to reorganize the structure of discordant datasets into one common format. It can be considered as a part of data preparation process that enables the final data consumers to analyse information in a consistent manner. There are three fundamental use case categories in dataset standardization:
\begin{enumerate}
    \item Elementary mapping from internal sources: manipulating internal datasets that originate from inconsistent definitions and reorganizing them into a single principled dataset for the entire organization;
    \item Elementary mapping from external sources: transforming data from systems external to the organization through mapping them to keys and values compatible with the organization schemas;
    \item Advanced reconciliation: creating sophisticated metrics that assure their own semantics grounded in defined business logic;
\end{enumerate}

In order to understand the essence of dataset standardization, one may imagine holding companies with a plethora of independent subsidiaries, franchisees or globally spread business units that provide inconsistent financial data, for instance through different formats of balance sheets. 

Furthermore, there is no uniformly agreed practice to present a dataset description, often leading to the miscommunication between the data engineers and the final data users. As a solution, several dataset description standardizations have been proposed in order to mitigate the negative consequences stemming from the misuse of data. Essentially, they define a specific schema to document the data collection process and final data composition. A non-exhaustive list includes:
\begin{enumerate}
    \item Datasheets for datasets \citep{bender_data_2018}: a framework designed primarily for the Natural Language Processing research;
    \item Data statements \citep{gebru_datasheets_2020}: a list of questions that each successful data documentation should address;
    \item Dataset nutrition labels \citep{holland_dataset_2018}: a module resembling the nutrition facts labels for packaged food;
    \item Global Standardization of Clinical Research Data \citep{global_standardization_2019}: a project to standardize data collection and sharing in clinical research;
    \item Codebook \citep{arslan_how_2019}: package to automate the data documentation, principally for the psychological datasets;
\end{enumerate}

Consequently, dataset standardization along with its description regularities contribute to the cleanliness of data landscape, hence improving team efficiency and reducing long-term maintenance costs.

\myparagraph{Exploratory data analysis (EDA)}
Assuming a consistent dataset structure, exploratory data analysis provides a summary of its primary characteristics, often with visual techniques. It includes both the extraction of statistical properties, such as dimensionality, missing observations or range, mean and variance, as well as a graphical visualization\footnote{The most often employed chart types can be easily accessed on \url{https://datavizcatalogue.com}}. Moreover, several operations on the data can be performed in order to prepare it for the machine learning algorithms, e.g.: median polishing, trimeaning or ordination. 

Additionally, since real-world datasets are generally high-dimensional, one can attempt to project them into a lower-dimensional representation while conserving its underlying characteristics to the largest extent possible.
Widely used dimensionality reduction techniques include Principal Component Analysis (PCA, recommended for a linear data structure),
t-distributed Stochastic Neighbour Embedding (t-SNE)
\footnote{Implemented in Python: \url{https://github.com/danaugrs/go-tsne}} 
or Latent Dirichlet allocation (LDA)
\footnote{Implemented in Python: \url{https://github.com/RaRe-Technologies/gensim}}.
Moreover, it is important not to overlook Local Linear Embedding \citep{roweis_nonlinear_2000})\footnote{Implemented in Python as part of scikit-learn library: \url{https://scikit-learn.org/stable/index.html}},
that consists of a nonlinear learning approach for generating low-dimensional neighbour-preserving representations from (unlabelled) high-dimension input and tends to be more precise in exploiting the underlying data structure than PCA.
Nevertheless, modern methods, such as Uniform Manifold Approximation and Projection \citep{mcinnes_umap:_2020} 
\footnote{Implemented in Python: \url{https://github.com/lmcinnes/umap}},
are constantly gaining momentum, mainly due to their scalability and thus computational speed.

Nonetheless, after the application of dimensionality reduction techniques, the ability to interpret the influence of individual features usually drops significantly. Therefore, it is essential not to overlook the rotations methods. They be decomposed into two main families:
\begin{enumerate}
    \item 1. Orthogonal, in which the factor variances get changed, but factors themselves remain uncorrelated and variable communalities are preserved:
    \begin{itemize}
        \item varimax: it aims to maximize variance among the squared values of loadings of each factor;
        \item quartimax: its objective is to minimize the number of factors needed to explain a variable and often produces the so-called general factor; 
    \end{itemize}
\item Oblique, in which factors are allowed to lose their uncorrelatedness if that will produce a clearer \textit{simple structure};
\end{enumerate}

EDA is primarily used for two reasons – to check the validity of the extracted data and to generate hypotheses and understanding on how to model the data at a later stage. To provide a concrete example, suppose the task is to model price elasticity of a selection of canned beverages. One of the ways EDA can bring value is through revealing if the analysed beverages have different volumes and as a consequence if one needs to control for the volume of liquid in the modelling, as depicted in the following chart.

\begin{figure}[h]
\captionsource{Distribution of volume of analyzed Stock Keeping Units (SKUs)}{Own elaboration}
\includegraphics[width=\textwidth]{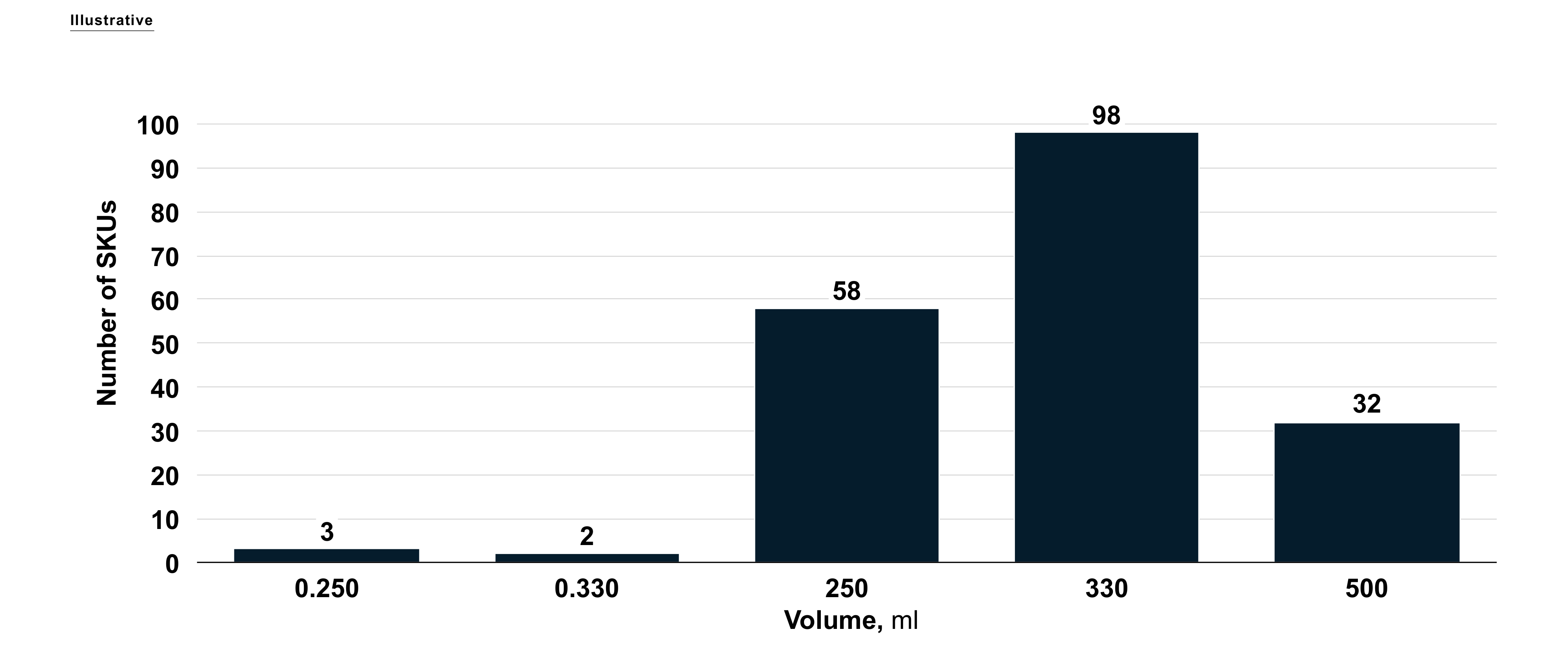}
\end{figure}

Furthermore, one might observe products with exceptionally low numbers, thus indicating that volume information in a single column is stored either in litres or millilitres and requires further investigation. In the outlined example, correcting units is relatively straightforward, since beverage cans have several well-known standard volumes, but typically expert domain knowledge is substantial.

Fortunately, manually performing EDA is becoming increasingly obsolete, as there is a plethora of available packages and toolkits for rapid and reliable extraction of the main properties from a dataset.\footnote{For a profound catalogue please feel free to visit \url{https://github.com/mstaniak/autoEDA-resources}}

\myparagraph{Dataset squashing}
Data squashing methods attempt to provide a subset of the training set which symbolizes the main essence of the dataset, that in turn can be granted along with the model prediction to the final user as the explanation, disentangling the need of storing the entire dataset. Usually, data summarization consists of selecting a minimal subset of representative samples (so called prototypes), for instance through the K-medoid clustering algorithm \citep{park_simple_2009}\footnote{Part of PyClustering library in Python: \url{https://github.com/annoviko/pyclustering}}.

Nevertheless, some argue that in order to completely understand complex data distributions, one need as well the parts of the input space that are not accurately described by the chosen set of prototypes (so called criticisms). Motivated by the Bayesian model criticism framework, \cite{NIPS2016_5680522b} developed MMD-critic technique (Maximum Mean Discrepancy)\footnote{Implemented in Python: \url{https://github.com/BeenKim/MMD-critic}} to efficiently learn both prototypes and criticisms.

The MMD-critic algorithm should be executed in two stages. Firstly, prototypes are elected to mimic the full dataset. Secondly, criticisms are selected from parts of the dataset that are under-represented by the previously chosen prototypes, with a supplementary constraint to assure the diversity of criticisms. As an output of MMD-critic method, a set of prototypes that are exemplary of the dataset as a whole, and a set of criticisms that diagnose spacious parts of the dataset that alter most from the prototypes are obtained. Titular Maximum Mean Discrepancy (MMD) refers to the specific scheme of quantifying of the difference between the prototype distribution and the full data distribution.

Additionally, Bayesian Case Model (BCM) proposed by \cite{kim2015bayesian}, extends the prototype-based reasoning to the fully-fledged and operational machine learning model. BCM is an admixture model, thus it belongs to the family of generative models for unsupervised learning), which leverages both prototypes and sparse features to be more interpretable (simply using examples to intuitively explain machine learning results to humans) without any loss of power compared to standard methods. It was further developed into an end-to-end interactive BCM (iBCM) system for the computer science education domain that communicates with humans through examples and accepts the feedback through identical intuitive medium.

Finally, it is valuable to outline the prominent and relatively mature data squashing techniques: moment-matching \citep{dumouchel_squashing_1999}, in which the squashed dataset consists of a set of pseudo data points selected to replicate the moments of the original dataset within subsets of a partition of the mother-data and likelihood-based methods \citep{madigan_likelihood_2002}, in which for every mother-data point the algorithm firstly estimates the probability of occurrence in order to generate a likelihood and then constructs one or more pseudo data points from each likelihood profiles clusters.\footnote{A helpful library to build a coreset (meaning the core of the dataset), for the purposes of Bayesian inference can be found here: \url{https://github.com/trevorcampbell/bayesian-coresets}}

Dataset summarization techniques can be employed in marketing campaign optimization, since it involves merging analytics with a business understanding to design the most effective communication. One of methods used by marketing specialists is to create personas, defined as a composite sketch of a key segment of targeted audience. Dataset summarization techniques can be therefore incorporated to extract the mixture of any of the attributes or activities that allow to define an accurate subset of representative personas instances. The chart below depicts selected characteristics of personas established in McKinsey “Digital Poles” survey \citep{digital_poles_mck}.

\begin{figure}[h]
\captionsource{Why do digital Poles still shop in traditional stores?}{Adapted from \cite{digital_poles_mck}}
\includegraphics[width=\textwidth]{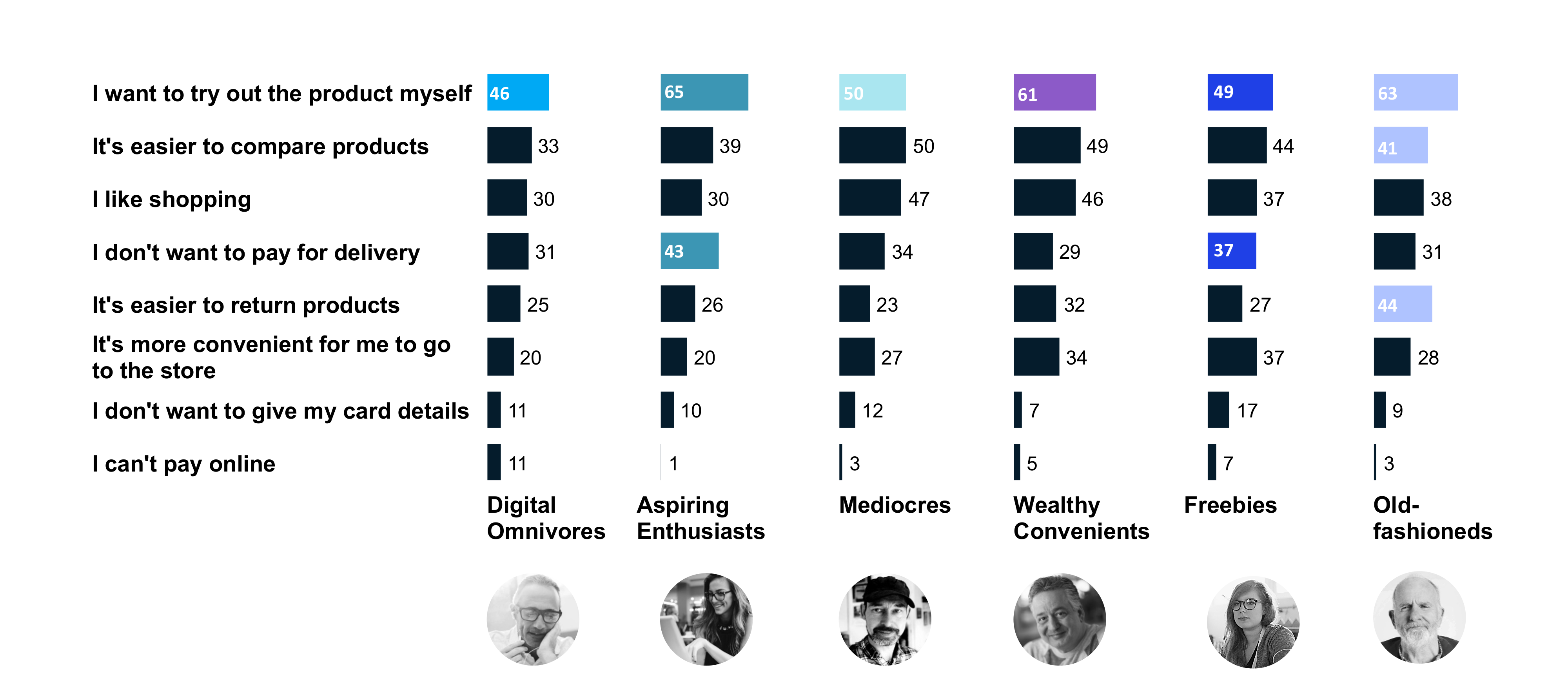}
\end{figure}

It presents different causes for shopping in traditional stores versus their e-commerce alternatives, constituting the basis for the growth potential of e-commerce market in Poland discussion. Product manager could consequently use them to tailor communication to different groups, for example: an e-commerce player emphasising low delivery costs will be more successful when aiming such marketing efforts at “Aspiring Enthusiasts” rather than “Wealthy Convenients”.

\myparagraph{Explainable feature engineering}
An explainable model should be based on the features that are easily interpretable. Thus, explainable feature engineering assures that the underlying bricks have been assigned an intuitive meaning. It can be tackled in two ways: domain-specific and model-based.

\textit{Domain specific} approach reckons on the expert knowledge, augmented with the insights derived from the exploratory data analysis, to identify valuable features. Circling back to the previous example of price elasticity modelling, a domain specific feature defined by data scientists could be the ratio of average prices of competing brands, which are known to be treated as substitutes by customers.

\textit{Model-based} approach, on the contrary, uncovers the underlying structure of a given dataset through mathematical models and transformations. The incessantly developing techniques include:
\begin{enumerate}
    \item 	Dictionary learning: it is designed to explore the input data in order to locate a sparse representation in the form of a linear combination of basic elements (so called dictionary). The procedure of dictionary learning can be therefore considered as a valuable alternative to Principal Component Analysis (PCA). Nevertheless, the atoms in the dictionary are not required to be orthogonal and additionally they may produce an over-complete spanning set, which allows for more flexible dictionaries and richer data representations. Consequently, the sparse dictionary learning has been successfully applied to various image, video and audio processing, as well as the unsupervised clustering tasks \citep{5494985}\footnote{Part of scikit-learn library in Python: \url{https://github.com/scikit-learn/scikit-learn}}.
    
	A generalization of k-means clustering method, K-SVD algorithm \citep{aharon_2006}, focuses on creating a dictionary for sparse representations via a singular value decomposition approach (SVD). More precisely, it iteratively modifies the sparse coding of the input data on the basis of current dictionary and consequently updates the atoms in the aforementioned dictionary in order to more precisely fit the data. Currently, K-SVD is widely applicable in biology or document analysis.
	\item Disentangled representation learning: a disentangled representation requires single latent units to be sensitive to modifications in single generative factors, while being invariant to changes in other. It allows the knowledge about one factor to relatively costlessly generalize to alternative configurations of other ones. For instance, a model trained on a dataset of 3D objects might learn independent latent units receptive to individual data generative factors, such as object position, scale or colour. In terms of specific algorithms, the extension of well-known variational autoencoder (VAE) framework \citep{kingma2014autoencoding}\footnote{Implemented with Keras in Python: \url{https://keras.io/examples/variational_autoencoder}}, $\mathbf{\beta}$ – VAE (Higgins et al., 2017)\footnote{Implemented with PyTorch in Python: \url{https://github.com/1Konny/Beta-VAE}}, tends to outperform the state of the art unsupervised \citep{chen2016infogan}\footnote{Implemented with Keras in Python: \url{https://github.com/tdeboissiere/DeepLearningImplementations}} and semi-supervised \citep{NIPS2015_ced556cd}\footnote{Implemented with Tensorflow in Python: \url{https://github.com/yselivonchyk/TensorFlow_DCIGN}} approaches to disentangled representative learning. In contrast to InfoGAN, $\mathbf{\beta}$ – VAE is stable to train, requires only few assumptions about the data and depends on tuning of a single hyperparameter, which in turn can be optimized through a hyperparameter search on the weakly labelled dataset or through heuristic visual inspection for purely unsupervised data.
\end{enumerate}

\subsubsection{Explainable modelling}

The following section is focused on the modelling stage of AI development, designed to avoid the black-box problem from the very beginning. In particular, it scrutinizes a handful of methodologies for training models that are both explainable by design and performant. Explainable modelling can be partitioned into two fundamental categories: implementation of inherently explainable model, augmented with the regularization, and the family of hybrid models, intending to merge white- and black-box approaches into one. The section is concluded with some additional Deep Learning specific solutions for explainable modelling.

\myparagraph{Implement inherently explainable model}
The classical approach to reach explainability is to implement a specific family of models that are regarded inherently explainable, such as the linear models, generalized additive models, Bayesian classifiers or decision trees. In other words, it consists of building models based on features that those who need to receive explanations inherently understand and maintain traceability of those features through the model. 

It has to be underlined that even in the family of inherently explainable models there is a spectrum between entirely transparent models, where one fully recognize how all the variables are jointly linked to each other, and models that are slightly constrained in their form, such as models that are imposed to inflate as one of the variables increases, or models that, ceteris paribus, favour features that domain experts have classified as significant \citep{rudin_stop_2019}. As an example of the latter, it is invaluable to mention Self-Explaining Neural Networks \citep{alvarezmelis2018robust}\footnote{Implemented in Python: \url{https://github.com/dmelis/SENN}}. It is a by-design approach that intends to produce stable explanations by assuming a specific structure for both model itself and its explainer. SENN designs self-explaining models in stages, progressively generalizing linear classifiers to complex yet architecturally explicit models.

One of the main use-cases for adopting explainable model families is to satisfy regulatory requirements of model transparency in industries such as banking or insurance. A common practice in insurance technical pricing\footnote{Technical pricing is modelling aimed at determining the expected value of claims that will be incurred by the insurer if they offer protection to a customer.} is constructing a generalized linear model with categorical variables only (continuous variables have to be binned beforehand). It can then be used as a trivial scoring card, allowing for direct tracing of every possible model prediction.

\begin{figure}[h]
\captionsource{Life insurance pricing model following the tariff cell archetype by fitting a generalized linear model on binned variables. Light blue shading points to tariff cells of an example customer.}{Own elaboration}
\includegraphics[width=\textwidth]{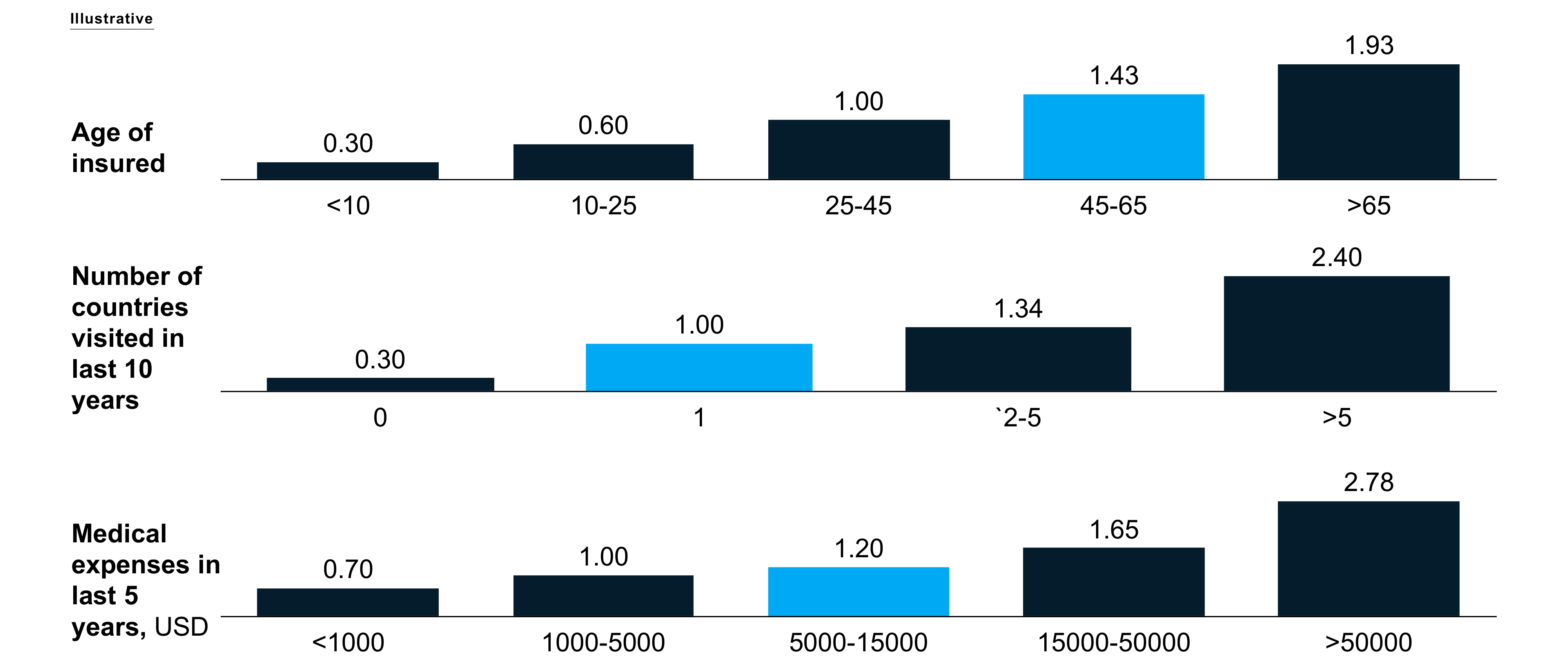}
\end{figure}

In the preceding example, one can easily trace each of the 100 possible tariff outcomes (so called tariff cells). The highlighted bars reflect a prospective customer who is between 45 and 65 years old, has visited 1 country in the last 10 years and has incurred a total of between 5,000 and 15,000 USD of medical expenses in the last 5 years. Technical price prediction can be then inferred by multiplying the relevant factors, i.e., $1.43 \cdot 1.00 \cdot 1.20 = 1.716$, meaning that the estimated technical price is 1.716 times higher than for a base case.

Unfortunately, explainable families generally result in relatively simple and poorly performing models, deficient in capturing the real-world complexity. Moreover, one needs to remember that simply using an explainable model does not mean the results of its processing will be fully explainable. For example, linear regression in a high-dimensional space is not \textit{transparent}, meaning that a person cannot contemplate the entire model at once, while even in the low-dimensional, yet multicollinear space, the estimated coefficients could be unstable. Fortunately, regularization can partially alleviate the aforementioned concerns.

\myparagraph{Regularization}
Generally speaking, regularization discourages learning a more complex or flexible model, in order to avoid the risk of overfitting. Consequently, it is heavily employed in order to improve the predictive performance of machine learning algorithms. However, its explainability boost characteristics are not to be underestimated. The basic implementations consist of explicitly restraining the model predictions to a specific value. More sophisticated ones attempt to incorporate regularization based on the expert knowledge, while the most advanced aim to deliver model agnostic regularizers not demanding any expert knowledge.

For instance, \cite{erion_improving_2020}\footnote{Implemented in Python: \url{https://github.com/suinleelab/attributionpriors}} developed a differentiable axiomatic feature attribution method called expected gradients that allow humans to use the common language of attributions to sanction prior expectations about a model's behaviour during training. For example, expected gradients can enforce models to treat functionally related genes similarly or to rely on fewer features for health care datasets. By straightforward encoding of constraints, the attribution priors tend to produce models with more intuitive behaviour and better generalization performance.

More advanced Contextual Decomposition Explanation Penalization \citep{rieger_interpretations_2020}\footnote{Implemented in Python: \url{https://github.com/laura-rieger/deep-explanation-penalization}} technique allows a practitioner to penalize both a model's prediction and the corresponding explanation, thus preventing a model from learning unwanted relationships and ultimately improving its predictive accuracy. To put differently, CDEP permits the user to directly penalize importance of certain features and their interactions, urging the neural network to not only deliver the proper predictions, but also the correct explanations for that predictions.  Moreover, CDEP is a relatively general approach, which can be applied to arbitrary neural network architectures, even on the sizeable datasets.

Additionally, \cite{plumb_regularizing_2019} proposed a strategy called Explanation-based Optimization (EXPO)\footnote{Implemented in Python: \url{https://github.com/GDPlumb/ExpO}} that permits to interpolate between inherently explainable models and post-hoc approaches by adding an interpretability regularizer to the loss function used to train the model. To be more specific, EXPO is a differentiable and model agnostic regularizer not requiring any domain knowledge, that approximates the fidelity metric
\footnote{This is a measure of explainer performance. For specific definition please visit evaluating XAI performance section.} on the training dataset in order to boost the excellence of post-hoc explanations of the model. It was shown empirically that EXPO can significantly improve explanation quality in terms of explanation fidelity, without sacrificing predictive accuracy.

There are three methods conceptually similar to EXPO and worth mentioning: Functional Transparency for Structured Data \citep{lee_functional_2019},
Right for the Right Reasons \citep{ross_right_2017}
\footnote{Implemented in Python: \url{https://github.com/dtak/rrr}} 
and Saliency Learning \citep{ghaeini_saliency_2019}. Nevertheless, FTDS lacks the model agnostic feature, since it is designed to graphs and time-series data, while RRR relies on extensive domain knowledge to specify the set of good and bad features.  To be more precise, RRR augments the loss function with the penalty term for the input gradients, primarily based on expert annotation, which constrains local explanations of a model to match the domain knowledge. Expert insights are encoded as a binary annotation matrix, indicating whether a specific feature should be taken into account while making a prediction for each input. Furthermore, when the annotations are not available, the validity of an explanation is assessed by the discovery of classifiers with similar accuracies, but qualitatively divergent decision boundaries for domain experts. Interestingly, the resemblance to the state-of-the-art sample-based explanation as Local Interpretable Model-Agnostic Explanations (LIME) has been confirmed empirically. Finally, recently emerged Saliency Learning targets to teach the model to undertake the right prediction for the relevant reason by providing explanation training and ensuring the alignment of the model's explanation with the ground truth explanation.

In addition, some authors aspire to approximate the results of the deep learning model by a modest decision tree\footnote{This is a post-modelling technique, but it sets context for the regularization method. For more examples of knowledge distillation please refer to post-hoc explainability section.}, which in turn can be easily tractable by humans (such model property is called human-simulatability, while the process of transferring the dark knowledge learned by a teacher model, usually sophisticated and large, to a student model, typically shallow and small, is named knowledge distillation). Benefiting from the existence of multiple local optima of a deep learning model, \cite{wu2017sparsity}\footnote{Implemented in Python: \url{https://github.com/dtak/tree-regularization-public}} presented that the aforementioned exercise can be sometimes achieved by supplementary regularization terms in the loss function. To be more precise, the novel regularization function penalizes the complexity of a decision tree (estimated by a meta-model as the average decision path length) approximating the initial model.

The last described regularization method has been widely used to create models for medical applications, namely prediction of mortality and need for ventilation in sepsis. Tree-regularized deep learning model allowed the researches to detect unwanted behaviour of the neural network by consulting with clinical experts – this would not have been possible without the enforced human-simulatable property.

As a remark, regularization can be employed additionally as a solution to vulnerability of deep neural networks to the adversarial attacks, meaning their susceptibility to errors after small modification of inputs. \cite{jakubovitz2019improving} depicted that the regularization using the Frobenius norm of the Jacobian of the network can lead to enhanced robustness results with a minimal change in the original network's accuracy.

\myparagraph{Hybrid models}
The purpose of hybrid approaches is to merge a model from an inherently explainable family with black-box techniques. The use of background knowledge in the form of logical statements or constraints in Knowledge Bases (KBs) has shown to not only improve explainability but also performance with respect to purely data-driven approaches \citep{garcez_neural-symbolic_2019}.
Future data fusion approaches may thus consider endowing Deep Learning algorithms with explainability by externalizing other domain information sources, for instance the interpretability inherent to probabilistic graphical models.
Hitherto, deep formulation of classical machine learning models has been employed for example in Deep Kalman Filters\footnote{Implemented in Python: \url{https://github.com/k920049/Deep-Kalman-Filter}}\citep{krishnan2015deep} or Structural Variational Autoencoders\footnote{Implemented in Python: \url{https://github.com/mattjj/svae}} \citep{NIPS2016_7d6044e9}.

A particular example of classical machine learning model enhanced with its deep learning counterpart was introduced by \cite{papernot_deep_2018} as Deep k-Nearest Neighbours (DkNN)\footnote{Implemented in Python: \url{https://github.com/Eric-Wallace/deep-knn}}.
Generally speaking, it is designed to combine the k-nearest neighbour algorithm with the representations of the data learned by each layer of the Deep Neural Network.

As far as image processing is concerned, \cite{wan_nbdt:_2021} proposed the Neural-Backed Decision Trees (NBDT)\footnote{Implemented in Python: \url{https://github.com/alvinwan/neural-backed-decision-trees}} technique to estimate a decision tree augmented with the information obtained from a simultaneous neural network. The process is as follows: firstly generate the induced hierarchy through hierarchical agglomerative clustering (e.g. classifying a Cat also as a Mammal and a Living Thing) and train the neural network  with a tree supervision loss (a cross entropy loss designed to separate representatives for internal nodes). Then, run inference by featuring images using the network backbone and run embedded decision rules.

Alternatively, one can consider using the BagNet architecture of \cite{brendel_approximating_2019}\footnote{Implemented in Python: \url{https://github.com/wielandbrendel/bag-of-local-features-models}}, which can be viewed as a bag-of-features (BoF) model, where the features are learned through the deep neural networks. In other words, BoF technique consists of a set of fundamental visual features and the number of their occurrences on each image (bag). For the sake of illustration, consider the task of classifying images into human and bird classes, on the basis of only two visual characteristics: a human eye and a feather. The simplest bag-of-features model increases the evidence for human for each eye in the image and similarly raise the chances for the bird for each feather. The class which accumulates the most evidence across the picture, will be the predicted one.

The idea behind BagNet architecture is then to merge the interpretability and transparency of BoF models with the excellent performance of Deep Neural Networks. The high-level strategy is following:
\begin{enumerate}
    \item Split the image into small image patches.
    \item Pass the aforementioned patches through a Deep Neural Network to obtain a class evidences for each patch.
    \item Sum the evidence over all patches to derive an image-level decision.
\end{enumerate}

Unfortunately, due to the task of image classification, the BagNet architecture tends to perform poorly when the local image features are not sufficient to learn the high-level characteristics.

In addition, \cite{wang_hybrid_2019} designed a general framework for hybrid predictive models. Their hypothesis is that there exists a subspace in the feature space where an uncomplicated and thus more understandable model can be almost as accurate as the black-box model. Therefore, as a substitute of explaining the black-box part, one can just supplant it on a subset of data where its employment is a pure overkill. Consequently, they designed a hybrid predictive model where input is employed by the interpretable model in the first place to verify if a prediction can be directly generated, while if not, the black-box model gets activated.

Moreover, at least in theory, a machine learning model can be trained to simultaneously provide a prediction and the corresponding explanation. For instance, \cite{hind_ted:_2019} proposed a Teaching Explanations for Decision (TED)\footnote{Part of AI Explainability 360 library in Python: \url{https://github.com/IBM/AIX360}} technique that, unlike previously mentioned methods, does not pursue to probe the reasoning proceeding of a model, but instead attempts to clone the reasoning mechanism of a human domain user.

Assuming that the training dataset composes of a set of features (connoted by X), a decision / label / classification for each feature vector (expressed as Y) and an explanation describing the rationale for that resolution (denoted by E), which can take any form, not to mention a number, text string or image. Dissimilar to the traditional approaches, E does not necessarily need to be expressed in terms of X. On the ground of the above-mentioned augmented training set, the TED framework encodes Y and E into a new prediction target (YE) by combining each label from Y with a relevant explanation from E, which together with X is provided as the final training input to any deep AI algorithm. The last step decodes the prediction from the form of YE into its individual components: Y and E.

For example, assume the goal is to predict whether a used car's price will increase in 10 years from now. Knowing the reason for a black-box model's prediction would certainly be helpful to a salesman trying to convince potential buyers that a prediction of increased future value is reasonable.

\begin{figure}[h]
\captionsource{Data used to predict change in used car's value before and after supplying explanations for features}{Own elaboration; visualization technique inspired by \cite{hind_ted:_2019}}
\includegraphics[width=\textwidth]{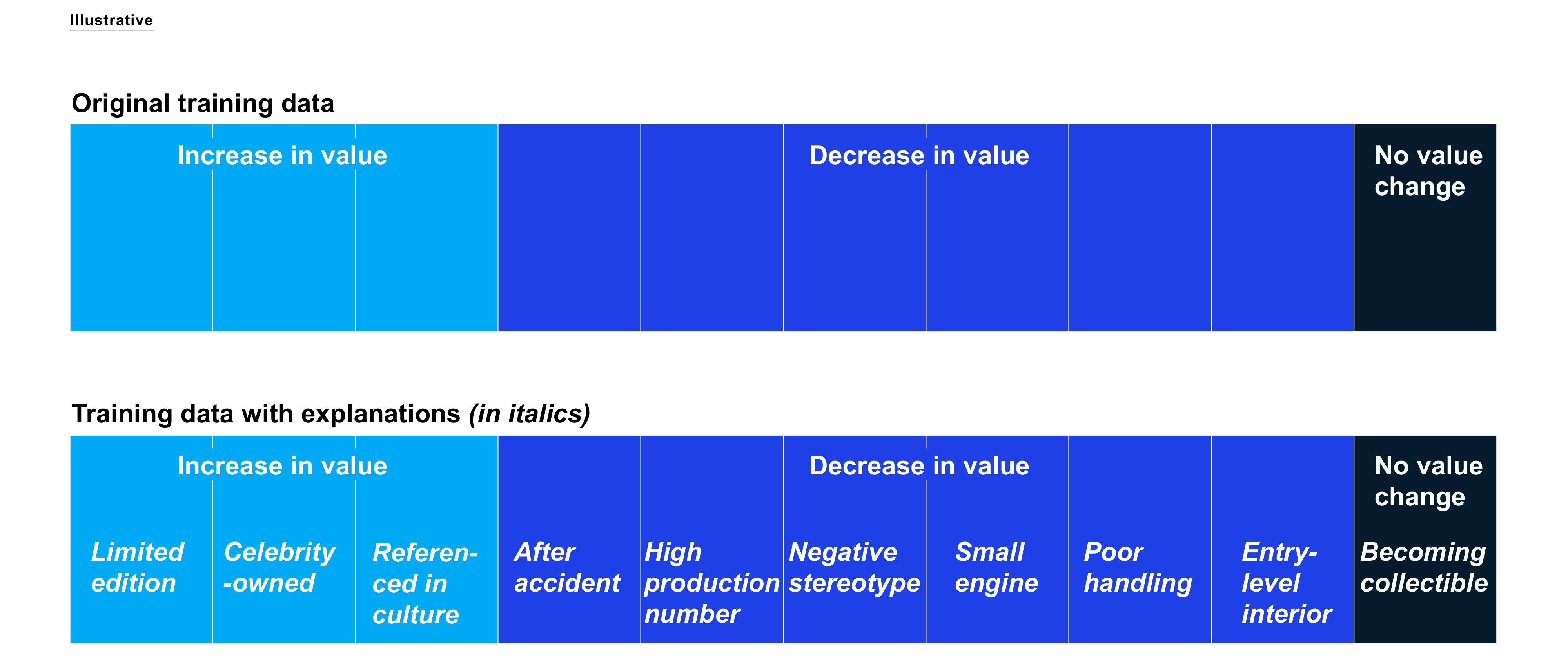}
\end{figure}

Some common examples of reasons why a car’s value might fluctuate are shown in the figure above. Using the TED framework, the model would learn to predict jointly the direction of value change and the underlying reason. Hence, a prediction of increase in value for a specific Porsche 911 could point the salesman that the expected surge in price stems from it being in fact a limited edition.

In a specific case of healthcare data, one can focus on the Reverse Time AttentIoN (RETAIN) model\footnote{Implemented in Python: \url{https://github.com/mp2893/retain}}, developed by \cite{choi2017retain}. It is founded on the two-level neural attention model that mimics the physician approach by digesting the data in a reverse time order, resulting in a recent clinical visit likely to receive higher attention. The attention mechanism helps in identifying the most meaningful visits and thus points out visit specific features that contribute to the overall prediction. However, it has to be emphasized that some experiments verified that the standard attention modules do not yet provide meaningful explanations and should not be treated as though they do \citep{jain_attention_2019}.

\myparagraph{Deep Learning specific solutions}
\cite{card_deep_2019} proposed the Deep Weight Averaging Classifiers (DWAC)\footnote{Implemented in Python: \url{https://github.com/dallascard/DWAC}} which is able to transform any deep learning architecture into one that is more transparent, interpretable, and robust to out-of-domain data. It eliminates the final SoftMax layer, responsible for the projection of a hidden representation into a vector of probabilities, replacing it by the probability of each label as a weighted sum of training instances, where the weights are based on the distance between instances in the low-dimensional instance-embedding space learned by the model.

Bayesian Deep Learning (BDL)\footnote{An example in Python: \url{https://github.com/kyle-dorman/bayesian-neural-network-blogpost}} gauges the uncertainty of a neural network about its predictions. It may model perplexing assignments by leveraging the hierarchical representation capabilities of deep learning, while additionally being able to infer sophisticated multi-modal posterior distributions. BDL models usually construct uncertainty estimates by either installing distributions over model weights, or by learning a direct mapping to probabilistic outcomes. By being acquainted with the weight distributions of sundry predictions and classes, one can tell which feature led to a specific decision and the relative importance of it in the uncertainty calibration.

Knowing the uncertainty of predictions is of vital importance in multiple applications ranging from safety-critical systems to optimization of expected profit from investments based on a model's predictions. BDL can be employed to model two crucial types of uncertainty: aleatoric and epistemic. 

Aleatoric uncertainty is stemming from the inability of the model to explain a particular data point. A simple example would be a performing computer vision on input from a camera, whose lens has been covered with snow, thus obstructing view of at least parts of the image. This type of uncertainty is especially relevant for systems optimized for safety, such as self-driving cars, where the correct decision could be to disengage the self-driving system when it is not aware of its surroundings to a satisfactory degree. 

Epistemic uncertainty, on the other hand, arises from a lack of sufficient training data translating to risk of incorrect prediction on data with features the model has not seen before. Trivially, this uncertainty can be reduced by increasing the size of the training set. However, in many use-cases this is not possible, e.g., in time series data epistemic uncertainty is inevitable as naturally information about the future is not available at prediction. Such example is provided by \cite{zhu_deep_2017}, who have employed Monte Carlo dropout\footnote{MC dropout is sometimes considered as post-modelling technique and thus more suitable for the next section.} to estimate a whole posterior distribution of the number of completed Uber trips, which in turn can be employed in anomaly detection. Monte Carlo dropout works by repeatedly and randomly deactivating neurons in the neural networks’ hidden layers during prediction\footnote{This should not be confused with a dropout layer in neural network architecture. MC dropout does not affect training of the network, it is used solely during inference.}. Consequently, each input is mapped to a vector of predictions, which can then be used to approximate model uncertainty by empirically calculating variance. Another possible use case is creating a framework for risk-adjusting profit expected from investments based on the uncertainty of prediction in a manner similar to Sharpe ratio.\footnote{For reference see \cite{sharpe_mutual_1966}.}

\subsubsection{Post-hoc explainability}

Post-modelling explainability aims to extract explanations in order to understand the pre-developed models. It can be partitioned into two fundamental categories: model-level and instance-level.

\myparagraph{Model-level post-modelling explainability}
Model-level post-hoc explainability aspires to provide understanding of model's prediction overall, for the whole set of observations. It usually starts with the assessment of statistical contribution of each feature to the estimated model. Variable importance calculation for linear and tree-based models is relatively straightforward, because it routinely exploits their notable structural characteristics, for instance through the determination of regression coefficients or computation of number of times a given feature is employed to split tree nodes.

Nevertheless, in order to enable the comparison of different model structures, one needs to operate with model-agnostic feature importance algorithms, such as the permutation-based version of \cite{altmann_permutation_2010} or its modern equivalent of \cite{fisher2019models}\footnote{Implemented in Python as part of scikit-learn library: \url{https://scikit-learn.org/stable/index.html}}, which measures the drop in model predictive performance when a single feature is randomly shuffled. One has to bear in mind that the aforementioned approach requires significant computational power for a model with a sizeable number of features, since it demands training a new model (on the perturbed data) for each predictor. Moreover, due to randomness in the shuffling step, it is advised to consider replicating it for several permutations and to ensure a lack of multicollinear or correlated features.

Another crucial subgroup of model-level techniques levies perturbations of explanation drivers, where the statistical contribution of a given predictor is assessed through the delta of model’s predictions after it is altered. Partial Dependence \citep{10.1214/aos/1013203451}\footnote{Implemented in Python as part of scikit-learn library: \url{https://scikit-learn.org/stable/index.html}} profiles depicts the value of model prediction as a function of a selected explanatory variable, typically through mean aggregation of individual Ceteris-Paribus (CP) profiles\footnote{CP profile illustrates the dependence of an instance-level prediction for a given feature, for more information please refer to instance-level post-modelling section.}. They can be helpful to determine whether the relationship between the target and a feature is linear, monotonic or more complex, as illustrated on the example below. 

\begin{figure}[h]
\captionsource{Partial dependence plot of predicted salary versus age}{Own elaboration}
\includegraphics[width=\textwidth]{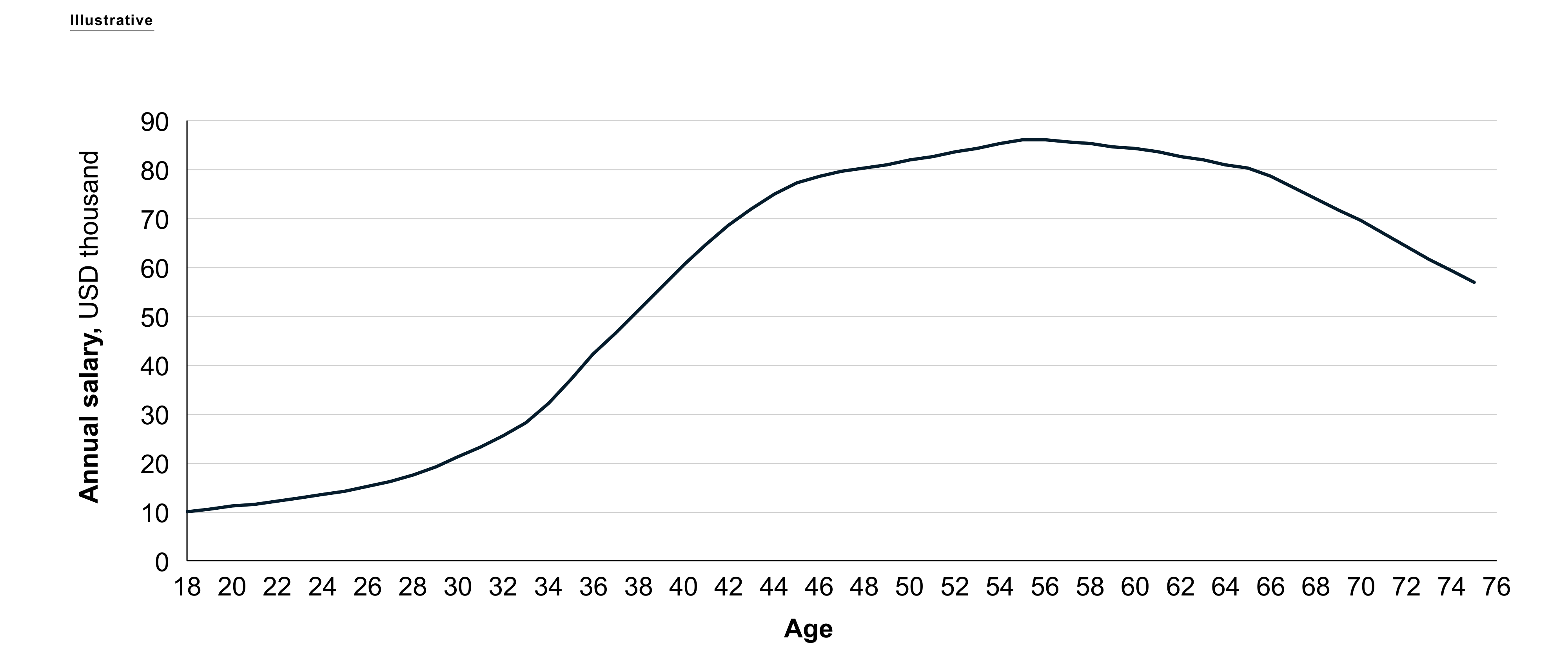}
\end{figure}

Recently proposed SHAP Dependence plot \citep{lundberg2017unified}\footnote{Implemented in Python as part of shap library: \url{https://github.com/slundberg/shap}} provides richer information than the traditional Partial Dependence plots, because it guarantees one additional piece of information, namely the variance of observations. In general, dependence plots can be easily compared across different model structures, nevertheless they may deliver misleading summarization in the case of correlated features. As a potential solution, one may consider Accumulated Local Effects (ALE) profiles, introduced by \cite{apley_visualizing_2019}\footnote{Implemented in Python: \url{https://github.com/blent-ai/ALEPython}}. When handling (even strongly) correlated variables, ALE employs actual conditional marginal distributions instead of considering each marginal distribution of features, therefore alleviating extrapolation of the response at predictor values that are far outside the multivariate envelope of the training data (required by PD).

Another indispensable group of model-level ex-post explainability techniques focuses on knowledge distillation, aiming to represent a sophisticated model behaviour as a set of decision rules. In case of networks with less than two layers, they can be further decomposed into three main categories. To start with, the decomposition algorithms attempt to split the network into neuron level and then the outcomes obtained from each neuron are aggregated to represent a network as a whole, for instance CRED technique of \cite{sato_rule_2001}.
Alternatively, pedagogical rule extraction algorithms treat a neural network as a black box, irrespective of its architecture, resulting in much translucent approach, for example TREPAN algorithm of \cite{Craven94usingsampling}.
However, since the decomposition technique works layer by layer, it might be tedious and time-consuming. Finally, the eclectics algorithms incorporate both the aforementioned approaches.

Unfortunately, all of the above-mentioned techniques are not especially applicable for Deep Learning. Hence,
\cite{10.1007/978-3-319-46307-0_29} suggested a novel rule extraction algorithm called DeepRED\footnote{Implemented in Python: \url{https://github.com/camgbus/DeepRED}}, by extending a decompositional CRED algorithm to arbitrary number of layers. The proposed algorithm has additional decision trees as well as intermediate rules for every hidden layer. Rule extraction from DNN is a divide and conquer process that characterizes each layer in terms of the previous layer, helping to reduce memory usage and computational time. Then, each result is merged to get the final rule that explains the whole deep neural network. Furthermore, \cite{liu_improving_2018} augmented the aforementioned approach to distil Deep Neural Networks into vanilla decision trees on multi-class datasets.

Unfortunately, although DeepRED can construct complete trees that are closely faithful to the original network, the generated trees can be large enough to make a model incomprehensible, and the implementation of the method still takes substantial time and memory, being therefore limited in scalability. Consequently, several approaches to learn the proxy models to imitate the behaviour of a complex model only locally have been developed. As an example, Black Box Explanations through Transparent Approximations (BETA) algorithm \citep{lakkaraju_interpretable_2017} learns a compact two-level decision set in which each rule explains part of the model behaviour unambiguously, reaching high fidelity (agreement between the explanation and the model), low imbrication (little overlaps between decision rules in the explanation), and high interpretability (the explanation decision set is lightweight and small).

Similarly, the study of \cite{henelius_interpreting_2017} introduced the Automatic STRucture IDentification (ASTRID)\footnote{Implemented in R: \url{https://github.com/bwrc/astrid-r}} technique to examine which attributes are exploited by a classifier in generating a prediction. This approach discovers the largest subset of features such that the accuracy of the classifier trained with it is indistinguishable in terms from a classifier built on the original feature set.

Some mechanisms produce the importance scores in terms of input features of the model, starting from the last layer, computing the contributions of its neurons to the target and then consecutively repeating the process layer by layer until the input layer is reached. The contributions from the previous layer can be estimated through partial gradients of a target neuron activation with respect to previous layer neurons as in the Guided Backprop\footnote{Implemented in Python \url{https://github.com/PAIR-code/saliency}} \citep{springenberg_striving_2015} and SmoothGrad\footnote{Implemented in Python: \url{https://github.com/PAIR-code/saliency}} \citep{smilkov_smoothgrad:_2017} techniques or by decomposing a target neuron activation value into its constituent values appearing from previous layer as in the Layer-wise Relevance Propagation\footnote{Implemented in Python: \url{https://github.com/sebastian-lapuschkin/lrp_toolbox}} \citep{bach_pixel-wise_2015} and Deep Learning Important FeaTures\footnote{Implemented in Python: \url{https://github.com/kundajelab/deeplift}} \citep{pmlr-v70-shrikumar17a} approaches.

Each technique adjusts a balance between highlighting areas of high network activation, where neurons impact strongest, and spaces of high network sensitivity, where modifications would mostly influence the output. A comparison of some of those can be found in \cite{ancona_2017}.

More mature, but still worth-mentioning approaches focused on rule extraction by direct mapping of inputs into outputs, rather than aiming to understand the inner workings of neural networks. In other words, they treated the model as a black box, but they searched for trends and functions of inputs to outputs. For instance, Validity Interval Analysis \citep{10.5555/2998687.2998750} mimics neural network behaviour in order to detect stable intervals, where the correlation between the input and the predicted class is present. Alternatively, the MofN algorithm \citep{towell_extracting_1993} locates the rules that explain single neurons by clustering and ignoring the insignificant neurons.

\myparagraph{Instance-level post-modelling explainability}
The second category of ex-post explainability provoke understandability of prediction for a single observation. The instance-level techniques usually impose perturbations of explanation drivers and then scrutinize and summarize their impact on a corresponding output. Therefore, the so-called perturbation mechanisms are model-agnostic and relatively straightforward to implement. However, the requirement of massive repetition of perturbations yields several impediments. Firstly, the construction of a meaningful perturbation is challenging. Ideally, the perturbations would be driven by the variation that is observed in the dataset. Thus, widely employed rule-of-thumb modifications, selected from a set of implausible alternative values can lead to dubious explanations. Secondly, a plausible interference should be conducted on a considerable amount of perturbed inputs, which results in a computational burden, particularly for models with high-dimensional datasets. Four main representatives of instance-level mechanisms are CP profiles, BD plots, LIME and SHAP.

Due to the similarity to the Partial Dependence (PD) plots from previous section, it is optimal to start with Ceteris Paribus (CP) profiles, sometimes referred to as Individual Conditional Expectations \citep{goldstein2014peeking}\footnote{Implemented in Python: \url{https://github.com/AustinRochford/PyCEbox}}. An ICE plot envisions the dependence of the prediction on a regressor for each instance separately, culminating in one line per observation, compared to one line overall in PD plots. The values for an aforementioned line can be calculated by creating synthetic variants of this specific instance by replacing the examined feature's value with values from a grid and computing accordingly predictions with the black-box model, keeping all other features unmodified (hence Ceteris Paribus). 

For models with a vast number of explanatory variables, the visual representations provided by ICE plots can be staggering. As a potential solution aiming to provide a summary among the set of features, one may consider Ceteris Paribus Oscillations \citep{biecek2021}\footnote{Part of DALEX package in R: \url{https://github.com/ModelOriented/DALEX}} that quantify the impact of a selected variable on model’s predictions. To be more specific, they can be viewed as the mean of the absolute deviations between the CP profile and the instance prediction.

Break-Down \citep{biecek2021}\footnote{Implemented in R: \url{https://github.com/pbiecek/breakDown}} plots should be considered as a valuable alternative to ICE plots. For each instance separately, they aim to select components of the input that cannot be modified without a significant delta in the prediction. To be more specific, for each regressor the model prediction is calculated with and without particular feature. Then, the feature contribution is computed as the distance to the initial model prediction\footnote{This approach can be therefore regarded as an approximation of Shapley values where each feature contribution is associated with the average effect of a feature across all feasible relaxations.}. After ordering, BS plots can be viewed as a model agnostic tool for decomposing the predictions from black boxes into a concise graphical way of a so-called waterfall chart.

\begin{figure}[h]
\captionsource{Break Down plot for ice cream sales prediction in a supermarket}{Own elaboration}
\includegraphics[width=\textwidth]{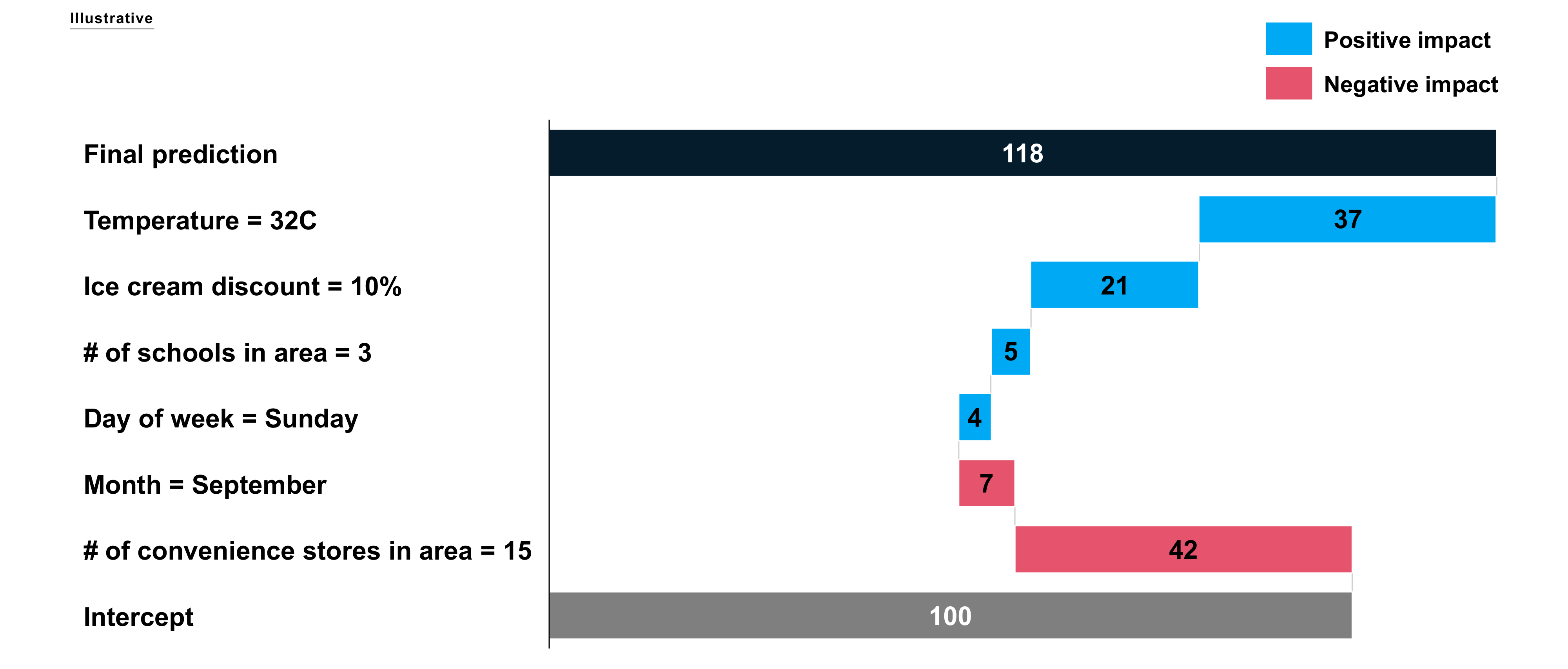}
\end{figure}

In the depicted example a BD plots of ice cream sales prediction has been provided. A store manager can then clearly see which factors are expected to impact demand positively and try to adjust the store accordingly.\footnote{Note the similarity in interpretation mechanism to the SHAP force plots introduced later.}

Local Interpretable Model-agnostic Explanations (LIME)\footnote{Implemented in Python: \url{https://github.com/marcotcr/lime}}, proposed by \cite{ribeiro_why_2016}, is a flagship example of the perturbation mechanism. In order to satisfy the model-agnostic property, LIME modifies the input to the model locally. In other words, instead of attempting to understand the entire model at once, an individual input instances are modified and their impact on the predictions monitored, typically by the approximation of the underlying model by an inherently interpretable one, such as a regularized linear model. The output of LIME is a list of explanations, reflecting the contribution of each feature to the prediction of a specific data observation, allowing to mitigate several biases through a human operator overriding the decisions upon evaluating the reasons given by the algorithm.

An invaluable alternative is the SHapley Additive exPlanations (SHAP) method\footnote{Implemented in Python: \url{https://github.com/slundberg/shap}}, derived by \cite{lundberg2017unified}.
The aforementioned framework exploits Shapley values, the notion from a cooperative game theory that measures the marginal contributions of players, to derive the importance of each input feature for a given instance prediction. Shapley values are obtained by averaging over every feasible sequence in which the feature could have been incorporated to the model. Alternatively, for cases where the feature is absent, it is represented as the expected value over the whole dataset. Naturally, sampling every possible feature subset of a long list of features is incredibly time-consuming, especially with high feature counts. Accordingly, a number of efficient Shapley value estimation approaches have been submitted for given model architectures, not to mention tree ensembles and deep networks.

SHAP can be particularly useful for situations in which a human agent needs to act upon particular prediction. A good example is uplift modelling, which is a technique employed in optimizing the efficiency of direct marketing campaigns. Suppose the goal is to help a bank's call centre in targeting customers that might be willing to buy life insurance through the bank's bancassurance partnership. Lead generation might be handled automatically by an uplift model created by the data science team, but materialization of impact heavily depends on operational execution. Conversely, the aforementioned execution is typically based on conversation scripts aiming to master call centre agents in maximizing conversion. Utilizing SHAP it is possible to elevate such scripts to a much higher sophistication level. 

\begin{figure}[h]
\captionsource{SHAP force plot depicting the impact of 5 variables on predicted uplift from marketing campaign}{Own elaboration}
\includegraphics[width=\textwidth]{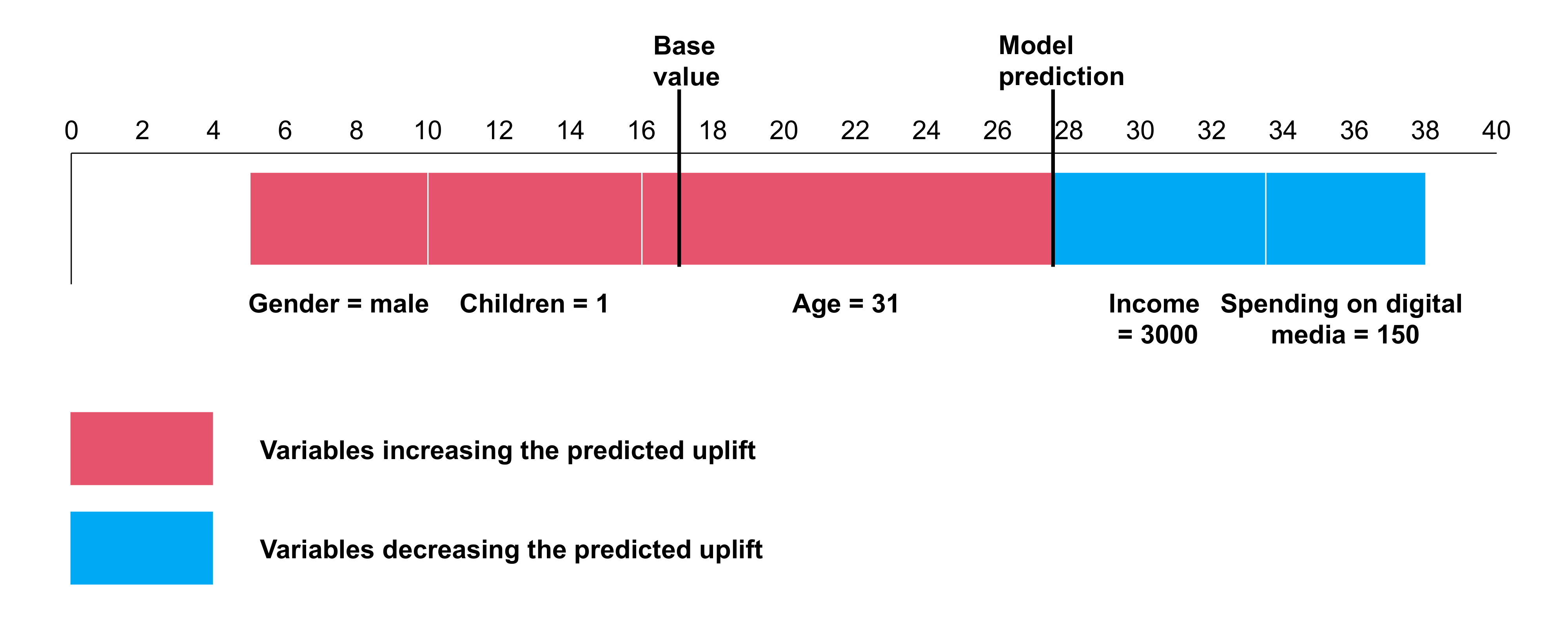}
\end{figure}

The SHAP force plot can be interpreted as follows: this particular lead is more likely than average customer to respond to the marketing campaign (positive model prediction). His likelihood to be persuaded to buy the life insurance is increased mainly by his age (equal to 31 years) and the fact that he is male with 1 child. On the contrary, his income of 3000 units and spending on digital media of 150 units discourage him to buy more than he would without being targeted. Consequently, SHAP force plot provides a crucial input to the agent calling the prospective customer – since age is the main driver of increased propensity to respond to the campaign, perhaps a script that emphasises entering a new stage in life will be the most useful in selling him the life insurance product. 

\subsubsection{Evaluating XAI Performance}

The opponents of the XAI algorithms often highlight their limitations, such as the lack of invariance to the simple transformations of the input (so called robustness of explanations, \cite{kindermans2017unreliability} or the singe point-wise explanation.
Moreover, it is not rare that two explainable artificial intelligence techniques provide two distinct interpretations.
In light of the aforementioned obstacles, \cite{gilpin_explaining_2018} enumerates additional types of measures quantifying the performance of XAI approaches:
\begin{enumerate}
    \item Completeness in comparison to the original model: an explanation model can be assessed directly according to how precisely it approximates the model being explained, also known as the fidelity.
    \item Completeness as quantified on a substitute task: an explanation model can be evaluated in terms of attributes different than direct model explanation, for instance a justification designed to reveal model sensitivity can be appraised against a brute-force measurement of the model sensitivity.
    \item Ability to detect biased models:  an explanation model can be gauged for its ability to expose models with the presence or absence of relevant biases.
    \item Human evaluation: an explanation model can be graded in terms of reasonableness, measuring the degree of a match between an explanation and human expectations. 
\end{enumerate}
The use of diverse metrics that align with the purpose and completeness of the targeted explanation is encouraged. In the rest of the section, the following measures are thoroughly described: the robustness of explanations and the fidelity of underlying classifier.

\myparagraph{Robustness of explanations}
\cite{alvarez-melis_robustness_2018} claim that a critical property that any interpretability method should satisfy is to be robust to local perturbations of the input, therefore, not to provide substantially different explanations for similar inputs. Simultaneously, it would guarantee that a simplified model could be approximately used in lieu of the true complex model, at least in a relatively small neighbourhood.

The first attempt to investigate the robustness is the visual inspection of attributes, however, it may be subjective and nearly infeasible for higher-dimensional inputs. Consequently, algebra literature proposes the notion of Lipschitz continuity, which measures the relative change in the output with respect to the input. To be more specific, because the typical definition of Lipschitz continuity is global, \cite{alvarez-melis_robustness_2018} suggest the generalization in the form of a point-wise, neighbourhood-based \textit{local Lipschitz continuity}, defined as:

\begin{definition}
Function $f:X\subseteq \mathbb{R}^n \to \mathbb{R}^m$ is \textbf{locally Lipschitz} if for every $x_0$ there exist $\delta > 0$ and $L\in \mathbb{R}$ such that $||x-x_0||<\delta$ implies $||f(x)-f(x_0)||\leq L||x-x_0||$ 
\end{definition}

In contrast to the global Lipschitz criterion, in the above-mentioned definition both $\delta$ and $L$ depend on the anchor point $x_0$. Moreover, the notion can be easily extended to handle models with discrete inputs as well. Unfortunately, the value of $L$ is rarely known a priori, instead it can be estimated through an optimization problem \citep[see]{alvarez-melis_robustness_2018}, for instance by the employment of Bayesian Optimization \citep{snoek2012practical}.

There is no universally desirable value of $L$, since its acceptable level depends on the application and the goal of interpretability. Therefore, it is useful to employ the robustness relatively across XAI techniques: an explainability algorithm with lower Lipschitz constant $L$ is more robust. Intuitively, more stable explanations (lower is better) tend to be more trustworthy \citep{ghorbani_interpretation_2019}.         

\myparagraph{Fidelity of the underlying classifier}
The interpretation framework is characterized by a high \textit{fidelity} if the prediction produced by an explanation eminently agrees with the original model \citep{plumb_regularizing_2019}.
Intuitively, an explainer with good fidelity precisely conveys which patterns the model has used in order to deliver a certain prediction. In other words, it depicts the degree to which each feature influences the model's prediction.

Moreover, satisfactory explanations should be able to illustrate a counterfactual dependence, illustrating the level of changes in explanandum (defined as the phenomenon to be explained) due to the modifications of initial conditions. As a consequence, the relevance of features can be measured as the impact of their changes to the value of explanandum. 

To be more specific, \cite{wachter_counterfactual_2017} argue that single predictions should be explained by w-counterfactuals that measure the minimum changes required for an observation to flip its classification.
A Counterfactual Local Explanations via Regression (CLEAR), introduced by \cite{white_measurable_2019}, grasp from that analysis, since it generates w-counterfactual explanations and then constructs local regression models. In contrast to the LIME method, which also builds regression to produce local explanations, in the next step the calculated w-counterfactuals are employed to improve the fidelity of its regressions.

\subsection{Expected evolution of XAI}

In recent years, progress of XAI has been promising, with efforts in explanation of deep learning processing yielding encouraging results.
However, there is an ample room for improvement, since diagnosing the predominant features and simplifying the problem space does not solve all possible obstacles associated with comprehending opaque models. 

In addition, XAI techniques should not overlook the format of explanations.
Bearing in mind that understandability is often provided to not data-savvy end-users, sophisticated statistical information provided by the vast majority of XAI algorithms can be overwhelming and consequently redundant.
Hence, presentation formats should be further enhanced to better promote user satisfaction. As a potential solution of more human-friendly explanations, one can focus on \textit{contrastive explanations}. Namely, instead of clarifying why a specific prediction was made, XAI can aim to explain why this prediction was made instead of another. For instance, consider a declined mortgage case. Then instead of answering the “Why have I not received the mortgage?”, maybe more plausible would be the response to “Why have I not received the mortgage, while my cousin has?”.

%\subsection{Additional resources}
%\begin{enumerate}
%    \item The most frequently employed XAI methodologies with a brief descriptions: \url{https://towardsdatascience.com/the-how-of-explainable-ai-pre-modelling-explainability-699150495fe4}.
%    \item Additional papers about interpretable AI, grouped into thematic sections: 
%    \begin{itemize}
%        \item \url{https://github.com/liupeng3425/interpretable-research}
%        \item \url{https://github.com/anguyen8/XAI-papers}
%        \item \url{https://github.com/pbiecek/xai\_resources}
%    \end{itemize}
%    \item Additional papers about fairness in AI, grouped into thematic sections: 
%    \begin{itemize}
%        \item \url{https://medium.com/fair-bytes/reading-list-for-fairness-in-ai-topics-337e8606fd8d}
%    \end{itemize}
%    \item Valuable books:
%    \begin{itemize}
%        \item Barocas, S., Hardt., M., Narayanan, A., (2019). Fairness and Machine Learning.
%        \item Biecek, P. and Burzykowski, T. (2019). Predictive models: Explore, explain, and debug.
%        \item Escalante, H. J., Escalera, S., Guyon, I., Baró, X., Güçlütürk, Y., Güçlü, U., \& van Gerven, M. (2018). Explainable and interpretable models in computer vision and machine learning. Springer.
%        \item Hall, P. and Gill, N. (2018). Introduction to Machine Learning Interpretability. O’Reilly Media, Incorporated.
%        \item Molnar, C. (2019). Interpretable machine learning. Lulu.com.
%        \item Samek, W. (2019). Explainable AI: Interpreting, explaining and visualizing deep learning. Springer Nature.
%        \item Zhou, J., \& Chen, F. (2018). Human and machine learning: Visible, explainable, trustworthy and transparent. Springer.
%    \end{itemize}
%\end{enumerate}

\section{Conclusion}
The paper has provided an overview of fairness and XAI to help the reader to understand the two concepts and their possible implications for business.

Since future implementations of AI in numerous fields will utilize methods from fairness and XAI in a similar manner as predictive performance scores are used today,  businesses should start preparing immediately.  Evaluating which of current use-cases
could benefit from fairness methods,  adopting relevant organizational practices to ensure interpretability and finally investing in knowledge
and tools to data practitioners can help transforming any currently lagging firm into the data-driven market leader.

\nocite{*}
\bibliographystyle{apalike}
\bibliography{main}

\end{document}